\documentclass[aps,showpacs,floatfix]{revtex4}

\thicklines
\begin{document}
\def\slash#1{#1 \hskip -0.5em / } 
\def\beq{\begin{equation}}
\def\eeq{\end{equation}}
\def\beqy{\begin{eqnarray}}
\def\eeqy{\end{eqnarray}}

\thispagestyle{empty}

\title{Polarization Observables for Two-Pion Production off the Nucleon}
%\vspace{1.cm}\\
\author{
W. Roberts$^{1,2}$\\
and \\
T. Oed$^1$
}
%}
\affiliation{$^1$Department of Physics, Old Dominion University, Norfolk, VA 23529,
USA \\
and \\
$^2$Continuous Electron Beam Accelerator Facility \\
12000 Jefferson Avenue, Newport News, VA 23606, USA.}

%\date{\today}
\begin{abstract}
We develop polarization observables for the processes $\gamma N\to\pi\pi N$ and $\pi N\to\pi\pi N$,
using both a helicity and hybrid helicity-transversity basis. Such observables are 
crucial if processes that produce final states consisting of a spin-1/2 baryon 
and two pseudoscalar mesons are to be
fully exploited for baryon spectroscopy. We derive relationships among the observables, as well as
inequalities that they must satisfy. We also discuss the observables that must be measured in
`complete' experiments, and briefly examine the prospects for measurement of some of these
observables in the near future.
\flushright{JLAB-THY-04-279}
\end{abstract}
%\end{titlepage}
%\newpage
\maketitle
\setcounter{page}{1}

\section{Introduction and Motivation}

Polarization asymmetries are an essential ingredient in the interpretation of
various meson production reactions in terms of the various resonances that
contribute to the processes as real or virtual intermediate states. For
instance, much of the information that we have on the light baryon resonances
has been garnered from pion-nucleon and kaon-nucleon scattering experiments. In
addition, photoproduction experiments have augmented the database of
measurements that provide information on these resonances. The
differential and total cross sections for these reactions, together with various
polarization observables, are used to extract the amplitudes for the process, and
these are then interpreted as arising from a number of resonant and
non-resonant contributions \cite{piona,kaona}.

For processes in which the final state consists of a nucleon (or a spin-1/2 baryon, in general) and
a pseudoscalar meson, the polarization observables, their relationship to helicity or transversity
amplitudes, and the measurements needed to extract each observable are all well documented
\cite{piona,kaona,storrow,walker,barker,bds}. For processes in which the final state contains two
pseudoscalar mesons (along with a nucleon), the state of development is much less complete. For the
most part, the final state with two pseudoscalar mesons and a nucleon (mainly $N\pi\pi$) has been
treated as arising from either of the quasi-two-body states $\Delta\pi$ or $N\rho$, followed by the
decay of the $\Delta$ or the $\rho$ \cite{delta}. The $N\rho$ channel in particular, or more
generally, the $NV$ channel, where $V$ is a vector meson, hase received some attention in recent
years \cite{tabakin}. A number of authors have formulated treatments based on more general
quasi-two-body approaches \cite{qtb}.

This approach has been reasonably successful in the past, as the available data
came from high energy experiments. With today's facilities running at all
energies from threshold up to relatively high energies, a more complete
description of polarization observables for the three-body final state such as
we have been describing is warranted. Indeed, such a description is essential in
order to fully exploit the high-precision data that will be forthcoming. It must
be stressed that experiments with more than a single pseudoscalar in the final
state have been touted as our best hope for finding the so-called missing
resonances \cite{missings}. It is therefore timely and crucial that the polarization observables
for such processes be elucidated in a more general framework, one that goes
beyond the quasi-two-body assumption.

The importance of polarization observables can not be overstated. In the case
of photoproduction of a single pseudoscalar meson, four complex amplitudes of
some sort -- helicity, transversity, or  Lorentz covariant, for example -- are
required to describe the process. Since one phase will always remain ambiguous,
this means that seven `numbers' are required at each kinematic point. The
differential cross section provides information only on the sum of the absolute
squares of these amplitudes. Polarization observables allow extraction of more
information, including phases. For production of two pseudoscalar mesons, the
same holds true. The process is described in terms of a number of amplitudes,
and the differential cross section, in the form of mass distributions, angular
distributions, or even five-fold differential distributions, only provides
information on the sum of the absolute squares of these amplitudes. This is
woefully inadequate for arriving at an unambiguous description of the process.
As with the processes in which a single pseudoscalar meon is produced, 
polarization information is crucial.

The rest of this article is organized as follows. For definiteness, we refer to
the final state that we treat as $N\pi\pi$, but the formalism we present is
valid for any final state that consists of a spin-1/2 baryon and two pseudoscalar
mesons. In addition, we also discuss final states with a single pseudoscalar
meson in the final state, for the sake of comparison and completeness. In the next section we
discuss the kinematics for the two- and three-body final states that we
consider. In section III we introduce the formalism and notation by discussing 
the processes $\pi N\to\pi N$ and $\pi N\to\pi\pi N$. In section IV, we turn our attention to the
processes $\gamma N\to\pi N$ and $\gamma N\to\pi\pi N$. In
section V, we discuss the prospects for measurements of some of these
observables at present facilities, especially JLab, as well as our conclusions.

\section{Kinematics and Cross Sections}

\subsection{Introduction}

For all of the processes we discuss, we work in the center of mass (com) frame, and
define the momentum of the beam particle (pion or photon) to be
\begin{equation}\label{phok}
k=\left(\omega,0,0,K\right)\equiv \left(\omega,{\bf K}\right),
\end{equation}
with the momentum of the target nucleon being
\begin{equation}\label{nucp}
p_1=\left(\sqrt{s}-\omega,0,0,-K\right).
\end{equation}
We define the collision plane in terms of the particle(s) recoiling
against the final nucleon. Then
\begin{equation}
q=\left(\sqrt{s}-E,p\sin{\theta},0,p\cos{\theta}\right),
\end{equation}
The $x-$, $y-$ and $z-$axes are then defined as
\begin{equation}
\hat{z}=\hat{K},\,\,\,\, 
\hat{y}=\frac{\widehat{\bf{ K}\times {\bf Q}}}{\left|{\bf K}\times{\bf Q}\right|},\,\,\,\, 
\hat{x}=\frac{\hat{y}\times\hat{z}}{\left|\hat{y}\times\hat{z}\right|}
\end{equation}
It is also useful to define a set of axes in which the $z^\prime-$axis is
parallel to the momentum of the recoil nucleon. In this system,
the $y$-axis coincides with the $y$-axis of the collision plane. In terms of
momenta, the axes for this system are
\begin{equation}
\hat{z}^\prime=\hat{P},\,\,\,\, 
\hat{y}^\prime=\frac{\widehat{\bf{ P}\times {\bf K}}}{\left|{\bf P}\times{\bf K}\right|},\,\,\,\, 
\hat{x}^\prime=\frac{\hat{y}^\prime\times\hat{z^\prime}}{\left|\hat{y}^\prime\times\hat{z}^\prime\right|}
\end{equation}

\subsection{$\pi N\to\pi N$}

In the com frame, we choose the momenta of the initial pion and
nucleon as in eqns. (\ref{phok}) and (\ref{nucp}), while those for the final pion and nucleon are 
\beq\label{kinepnpn}
q=(\omega,Q\sin{\theta},0,Q\cos{\theta}),\,\,\,\,
p_2=(\sqrt{s}-\omega,-Q\sin{\theta},0,-Q\cos{\theta}),\,\,\,\,Q=K
\eeq
and $\sqrt{s}$ is the total center of mass energy. The Mandelstam 
variables of interest are
\begin{eqnarray}
s&=&(k+p_1)^2=(q+p_2)^2,\nonumber\\
t&=&(q-k)^2=(p_1-p_2)^2=2m^2-2(\sqrt{s}-\omega)^2-2K^2\cos{\theta},
\end{eqnarray}
where $m$ is the nucleon mass. The energies and momenta are
\begin{eqnarray}\label{sing3}
\omega&=&\frac{s+\mu^2-m^2}{2\sqrt{s}},\nonumber\\
Q&=&K=\lambda^{1/2}(s,m^2,\mu^2)/(2\sqrt{s}),\\
\end{eqnarray}
where $\lambda(a,b,c)=a^2+b^2+c^2-2ab-2bc-2ac$ is K$\not\!a$llen's function, and $\mu$
is the pion mass.

The cross section for the process is 
\beq
d\sigma=\frac{\overline{|{\cal M}_{fi}|^2}}{4\sqrt{(p_1.k)^2-m^2\mu^2}}
(2\pi)^4\delta^4(p_1+k-p_2-q)\frac{d^3p_2}{(2\pi)^32E} \frac{d^3q}
{(2\pi)^32\omega}.\nonumber
\end{equation}
After integration, this yields
\begin{eqnarray}
d\sigma=\frac{\,\overline{|{\cal M}_{fi}|^2}d\Omega_2}{16(2\pi)^2}.\nonumber
\end{eqnarray}

\subsection{$\gamma N\to\pi N$}

In the com frame, for real photons, the beam and target momenta are again as in eqns. (\ref{phok}) and
(\ref{nucp}), but now $K=\omega$.
For the pion and nucleon in the final state, the momenta are 
\beq\label{kinegnpn}
q=(\omega^\prime,Q\sin{\theta},0,Q\cos{\theta}),\,\,\,\,
p_2=(\sqrt{s}-\omega^\prime,-Q\sin{\theta},0,-Q\cos{\theta}).
\eeq
The Mandelstam variable $s$ has the same definition as before, but $t$ now takes the form
\beq
t=(q-k)^2=(p_1-p_2)^2=2m^2-2(\sqrt{s}-\omega)(\sqrt{s}-\omega^\prime)-2QK\cos{\theta},
\eeq
\begin{eqnarray}
K=\omega&=&\frac{s-m^2}{2\sqrt{s}},\,\,\,\, \omega^\prime=\frac{s+\mu^2-m^2}{2\sqrt{s}},
\,\,\,\,Q=\lambda^{1/2}(s,m^2,\mu^2)/(2\sqrt{s}).\nonumber
\end{eqnarray}
After phase space integrations, the cross section for this process is 
\begin{eqnarray}
d\sigma=\frac{Q\,\overline{|{\cal M}_{fi}|^2}d\Omega_2}{16(2\pi)^2\sqrt{s}K}.\nonumber
\end{eqnarray}

\subsection{$\pi N\to\pi\pi N$}

For this process, $k$, $p_1$ and $p_2$ are defined as in the $\pi N\to\pi N$ 
process. $q_1$ and $q_2$
are the momenta of the two final state pions, and momentum conservation gives
\beq
p_1+k=p_2+q_1+q_2.
\eeq
The momentum of the recoiling nucleon is taken to be
\begin{equation}\label{kinepnppn1}
p_2=\left(\sqrt{s}-s_{\pi\pi},-Q\sin{\theta},0,-Q\cos{\theta}\right),
\end{equation}
where 
\begin{equation}
s_{\pi\pi}=\left(q_1+q_2\right)^2
\end{equation}
and
\begin{equation}
Q=\frac{\lambda^{1/2}\left(s,s_{\pi\pi},m^2\right)}{2\sqrt{s}}.
\end{equation}
Here, we are using the recoiling nucleon, or more precisely, the recoiling pair of pions, to define the
collision plane. The momentum of one of the pions may be written
\beq\label{kinepnppn2}
\vec{q}_1=Q_1\left(\sin{\theta_1}\cos{\phi_1},\sin{\theta_1}\sin{\phi_1},\cos{\theta_1}\right),
\eeq
where $Q_1$, $\theta_1$ and $\phi_1$ can be written in terms of $s$,
$s_{\pi\pi}$, $\theta$, and the angles describing the motion of the pair of
pions in their cm frame. The expressions are too complicated to be reproduced
here.

We define the Mandelstam variables $s$ and $t$ as
\begin{equation} 
s=(k+p_1)^2,\,\,\,\, t=(p_1-p_2)^2.
\end{equation}
In addition, we may define a number of other Mandelstam variables as
\beqy
s_{N\pi_1}&=&(p_2+q_1)^2,\,\,\, \, s_{N\pi_2}=(p_2+q_2)^2,\nonumber \\
t_1&=&(k-q_1)^2,\ t_2=(k-q_2)^2.
\eeqy
Note that $s_{\pi\pi}$, $s_{N\pi_1}$ and $s_{N\pi_2}$ are not all independent, as they
satisfy
\beq
s_{\pi\pi}+s_{N\pi_1}+s_{N\pi_2}=s-2m^2-2\mu^2
\eeq

The differential cross section for this process is described in terms of 5 
kinematic variables. These may be, for instance, two Lorentz invariants and 
three angles. One obvious choice for one of the invariants is $s$. The choice 
of the other four quantities can be fairly arbitrary, and will depend on what 
information is being presented.  One choice is the scattering angle of the 
nucleon, $\theta$, or equivalently, $t$. For the other three variables, we can 
choose, for example, $s_{\pi\pi}$ and 
$d\Omega^\star_{\pi\pi}\equiv d\theta^\star_{\pi\pi} d\phi^\star_{\pi\pi}$, consistent with the way we
define the momenta. Another equally valid choice would be 
$s_{N\pi_1}$ and $d\Omega^\star_{N\pi_1}$, where the solid 
angle is defined in the rest frame of the nucleon-pion pair. In this work,
we choose the kinematic variables $s$, $t$, $s_{\pi\pi}$ and 
$d\Omega^\star_{\pi\pi}$. 

The differential cross section is
\beqy
d\sigma&=&\frac{\overline{\left|{\cal
M}\right|^2}}{4\sqrt{(k.p_1)^2-m^2\mu^2}}\nonumber\\
&&\times(2\pi)^4\delta^4(p_1+k-p_2-q_1-q_2) \frac{d^3p_2}{(2\pi)^3 2E_2}
\frac{d^3q_1}{(2\pi)^3 2\omega_1}\frac{d^3q_2}{(2\pi)^3 2\omega_2},
\eeqy
where ${\cal M}$ is the amplitude for the transition, $E_2$ is the energy of the
recoiling nucleon, and $\omega_i$ is the energy of the $i$th pion.
Carrying out the integrations using standard techniques yields
\begin{equation}
\frac{d^5\sigma}{ds_{\pi\pi} d\Omega_{\pi\pi}^\star d\cos{\theta}}=\frac{1}{4}
\frac{1}{128 (2\pi)^4s^{3/2}K\sqrt{s_{\pi\pi}}}\overline{\left|\cal
M\right|^2}\lambda^\frac{1}{2}( s_{\pi\pi},\mu^2,\mu^2) 
 \lambda^\frac{1}{2}(s,s_{\pi\pi},m^2), 
\end{equation}
where $4 \mu^2\leq s_{\pi\pi} \leq (s-m)^2$, and
$K=\lambda^{1/2}(s,m^2,\mu^2)/(2\sqrt{s})$.   

\subsection{$\gamma N\to\pi\pi N$}

The kinematic treatment of this process is very much the same as for the
process $\pi N\to\pi\pi N$. The main difference arises in the fact that, for
the initial photon, $k^2=0$. We therefore do not discuss the kinematics of this
process any further at this point, except to write down the form for the
differential cross section. This is
\begin{equation}
\frac{d^5\sigma}{ds_{\pi\pi} d\Omega_{\pi\pi}^\star d\cos{\theta}}=\frac{1}{4}
\frac{1}{128 (2\pi)^4(s-m^2)s\sqrt{s_{\pi\pi}}}\overline{\left|\cal
M\right|^2}\lambda^\frac{1}{2}( s_{\pi\pi},\mu^2,\mu^2) 
 \lambda^\frac{1}{2}(s,s_{\pi\pi},m^2).
\end{equation}

\section{Observables in $\pi N \to \pi N$ and $\pi N \to\pi\pi N$}

For the processes $\pi N \to \pi N$ and $\pi N\to\pi\pi N$, the matrix element
${\cal M}$ can be written
\beq
i{\cal M}=\chi^\dag\left({\cal A}+ \vec{\sigma}\cdot\vec{\cal B}\right)\phi,
\eeq
where $\chi$ and $\phi$ are the Pauli spinors representing the final and initial
nucleons, respectively. Here, the quantity ${\cal A}+ \vec{\sigma}\cdot\vec{\cal
B}$ is the most general $2\times 2$ matrix that can be constructed, and 
${\cal A}$ and $\vec{\cal B}$ are quantities that will
contain all of the details of the `model' used to describe the particular
reaction being considered. At this point, their exact form is of no consequence.
For both processes, we choose the nucleon momenta as defined in eqns.
(\ref{nucp}) and (\ref{kinepnpn}).

These two processes may be described in either the helicity or
transversity basis. In the helicity basis, the axis of quantization of the
spin of each nucleon is its direction of motion. For the initial nucleon, the 
helicity spinors are,
\begin{equation} \label{eq:helicityspin1/2}
\phi^+=\left(\begin{array}{c} 0 \\ 1\end{array} \right),\,\,\,\,
\phi^-=\left(\begin{array}{c} 1 \\ 0\end{array} \right),
\end{equation}
while for the final nucleon, they are,
\begin{equation}\label{helicity}
\chi^+=\left(\begin{array}{c} - \sin\frac{\theta}{2} \\ 
\cos\frac{\theta}{2}\end{array} \right),\,\,\,\,
\chi^-=\left(\begin{array}{c} \cos\frac{\theta}{2} \\ 
\sin\frac{\theta}{2}\end{array} \right).
\end{equation}

In the transversity basis, the axis of quantization of the spin of each nucleon
is the $y$-axis, which is as previously defined. For either initial or final 
nucleon, the transversity spinors are 
\begin{equation}\label{transverse}
\phi^+_T=\frac{1}{\sqrt{2}}\left(\begin{array}{c} -i \\ 1\end{array} \right),\,\,\,\,
\phi^-_T=\frac{1}{\sqrt{2}}\left(\begin{array}{c} i \\ 1\end{array} \right),
\end{equation}
where the $\pm$ refer to the spin projection relative to the $y$-axis. 

For either of the processes being discussed,
\beq
{\cal A}+ \vec{\sigma}\cdot\vec{\cal B}=\left(\begin{array}{cc} 
{\cal A}+{\cal B}_3& {\cal B}_-\\
{\cal B}_+& {\cal A}-{\cal B}_3\end{array}\right)\equiv\left(\begin{array}{cc} A_+& B_-\\
B_+& A_-\end{array}\right),
\eeq
where we have defined
\beq
A_\pm\equiv {\cal A}\pm {\cal B}_3, \,\,\,\, B_\pm\equiv {\cal B}_1\pm 
i{\cal B}_2,
\eeq
and the ${\cal B}_i$ are the Cartesian components of $\vec{\cal B}$.

In terms of these, the four possible helicity amplitudes, $i{\cal
M}_{\lambda_N\lambda_N^\prime}$, are
\beqy
i{\cal M}_{++}&=&A_- \cos{\frac{\theta}{2}}-B_-\sin{\frac{\theta}{2}}\equiv {\cal M}_1,\nonumber\\
i{\cal M}_{+-}&=&A_- \sin{\frac{\theta}{2}}-B_+\cos{\frac{\theta}{2}}\equiv {\cal M}_2,\nonumber\\
i{\cal M}_{-+}&=&-A_+ \sin{\frac{\theta}{2}}+B_+\cos{\frac{\theta}{2}}\equiv {\cal M}_3,\nonumber\\
i{\cal M}_{--}&=&A_+ \cos{\frac{\theta}{2}}+B_+\sin{\frac{\theta}{2}}\equiv {\cal M}_4.
\eeqy
In these equations, $\lambda_N$ is the helicity of the target nucleon, while
$\lambda_{N^\prime}$ is that of the recoil nucleon. Note that in the form written
above, $A_-$ and $B_-$ occur in one block of helicity amplitudes, while $A_+$
and $B_+$ occur in another block, with no `mixing' between the blocks. This makes inverting the
equations very easy, yielding
\beqy 
{\cal A}&=&\frac{1}{2}\left[\left({\cal M}_1+{\cal M}_4\right)\cos{\frac{\theta}{2}}+
\left({\cal M}_2-{\cal M}_3\right)\sin{\frac{\theta}{2}}\right],\nonumber\\
{\cal B}_1&=&\frac{1}{2}\left[\left({\cal M}_2+{\cal M}_3\right)\cos{\frac{\theta}{2}}+
\left({\cal M}_4-{\cal M}_1\right)\sin{\frac{\theta}{2}}\right],\nonumber\\
{\cal B}_2&=&\frac{i}{2}\left[\left({\cal M}_2-{\cal M}_3\right)\cos{\frac{\theta}{2}}-
\left({\cal M}_1+{\cal M}_4\right)\sin{\frac{\theta}{2}}\right],\nonumber\\
{\cal B}_3&=&\frac{1}{2}\left[\left({\cal M}_4-{\cal M}_1\right)\cos{\frac{\theta}{2}}-
\left({\cal M}_2+{\cal M}_3\right)\sin{\frac{\theta}{2}}\right].
\eeqy
At this point, we have not yet used the parity properties of the helicity amplitudes. This will
be discussed later.

Two sets of transversity amplitudes may be defined. The first set are
obtained by direct application of the transversity spinors defined above. We
define these to be $i b_{\tau_N\tau_N^\prime}$, where $\tau_i=\pm$ is the 
projection of the spin of the state $i$ along the $y$-axis (with the 1/2 
suppressed), and these take the form
\beqy
i b_{++}&=&{\cal A}+ {\cal B}_2\equiv b_1,\nonumber\\
i b_{+-}&=&-\left({\cal B}_3+i{\cal B}_1\right)\equiv  b_2,\nonumber\\
i b_{-+}&=&-\left({\cal B}_3-i{\cal B}_1\right)\equiv  b_3,\nonumber\\
i b_{--}&=&{\cal A}- {\cal B}_2\equiv  b_4.
\eeqy
The `block' structure is again apparent, and inverting these gives
\beqy
{\cal A}&=&\frac{1}{2}\left(b_1+b_4\right),\,\,\,{\cal B}_1=\frac{i}{2}\left(b_2-b_3\right),\nonumber\\
{\cal B}_2&=&\frac{1}{2}\left(b_1-b_4\right),\,\,\,{\cal B}_3=-\frac{1}{2}\left(b_2+b_3\right).\nonumber
\eeqy
For observables defined in terms of these transversity amplitudes, the
$x^\prime$, $y^\prime$ and $z^\prime$ axes that define the Cartesian components
of polarization observables coincide with the axes that define the initial state. This
is because the transversity spinors contain no explicit information about the 
angles defining the motion of the recoil nucleon.

We can write the reaction rate $I$, as
\beq
\rho_f I=I_0\left[1+\vec{\Lambda}_i\cdot\vec{P}+\vec{\sigma}\cdot\vec{P}^\prime+
\Lambda_i^\alpha\sigma^{\beta^\prime}{\cal O}_{\alpha\beta^\prime}\right]
\eeq
where $\vec{P}$ represents the polarization asymmetry that arises if the target nucleon is
polarized, $\rho_f=\frac{1}{2}\left(1+\vec{\sigma}\cdot \vec{P}^\prime\right)$ is the density matrix of the recoiling nucleon, and ${\cal
O}_{\alpha\beta^\prime}$ is the observable if the Cartesian $\alpha$ component of the target polarization is 
known and the $\beta^\prime$ component of the recoil polarization is measured.
The primes indicate that the recoil observables, defined in terms of helicity
amplitudes, are measured with respect to the set of axes
$x^\prime,\,y^\prime,\,z^\prime$, previously defined. If the observables are
defined in terms of transversity amplitudes, the $x^\prime,\,y^\prime,
\,z^\prime$ axes are the same as the $x,\,y, \,z$ axes. $\vec{\Lambda}_i$ is the polarization 
of the initial nucleon.

\begin{table}[h]
\caption{Polarization observables in single and double pion production using a 
pion beam, expressed in terms of helicity and transversity amplitudes. 
Variables labeled with a T require a polarized target with recoil polarization
unmeasured, while those labeled with an R require an unpolarized target, but
the recoil polarization is measured. Those denoted TR require polarized
targets, with recoil polarization measured. The measurements required are shown
by the pair $\left\{t,r\right\}$, which denote the component of the target (t)
or  recoil (r) polarization that must known or measured. For the target
polarization, the $x$-, $y$- and $z$-axes are as defined in the text. The 
$x^\prime$-, $y^\prime$- and $z^\prime$-axes are also defined in the text, as is
the notation for the transversity amplitudes.}
\label{tab:singlepolobspi}
\begin{center}
\begin{tabular}{c|l|l|c|c}
\hline 
Observable & Helicity Form & Transversity Form & Expt. Required & Type\\
\hline 
$I_0$ & $ \left|{\cal M}_1\right|^2 + \left|{\cal M}_2\right|^2 + 
\left|{\cal M}_3\right|^2 + \left|{\cal M}_4\right|^2$
 & $ \left|b_1\right|^2 + \left|b_2\right|^2 + \left|b_3\right|^2 + 
\left|b_4\right|^2 $ & $\left\{-;-\right\}$ & \\[+10pt] 
$I_0P_x$ & $ 2\Re\left({\cal M}_1{\cal M}_3^* + 
     {\cal M}_2{\cal M}_4^*\right) $
 & $ -2\Im\left(b_1b_3^* +b_2b_4^*\right) $ & $\left\{x;-\right\}$ & T\\
$I_0P_y$ & $ -2\Im\left({\cal M}_1{\cal M}_3^* + 
      {\cal M}_2{\cal M}_4^*\right) $ 
       & $ \left|b_1\right|^2 + \left|b_2\right|^2 - \left|b_3\right|^2 - \left|b_4\right|^2 $ & $\left\{y;-\right\}$ & \\
$I_0P_z$ & $ -\left|{\cal M}_1\right|^2 - 
     \left|{\cal M}_2\right|^2 + \left|{\cal M}_3\right|^2 + 
     \left|{\cal M}_4\right|^2 
$ & $ -2\Re\left(b_1b_3^* + b_2b_4^*\right) $ & $\left\{z;-\right\}$ & \\[+10pt] 
$I_0P_{x^\prime}$ & $ -2\Re\left({\cal M}_1{\cal M}_2^* +
     {\cal M}_3{\cal M}_4^*\right) 
$ & $ 2\Im\left(b_1b_2^* +b_3b_4^*\right) $ & $\left\{-;x^\prime\right\}$ & R\\
$I_0P_{y^\prime}$ & $ 2\Im\left({\cal M}_1{\cal M}_2^* + 
      {\cal M}_3{\cal M}_4^*\right) 
$ & $\left|b_1\right|^2 - \left|b_2\right|^2 + \left|b_3\right|^2 - \left|b_4\right|^2 $& 
$\left\{-;y^\prime\right\}$ &  \\ 
$I_0P_{z^\prime}$ & $ \left|{\cal M}_1\right|^2 - 
     \left|{\cal M}_2\right|^2 + \left|{\cal M}_3\right|^2 - 
     \left|{\cal M}_4\right|^2 
$ & $ -2\Re\left(b_1b_2^* + b_3b_4^*\right) $ & $\left\{-;z^\prime\right\}$ & \\[+10pt] 
$I_0{\cal O}_{xx^\prime}$ & $ -2\Re\left({\cal M}_2{\cal M}_3^* +  
     {\cal M}_1{\cal M}_4^*\right) 
$ & $ 2\Re\left(-b_2b_3^* + b_1b_4^*\right) $& $\left\{x;x^\prime\right\}$ & TR \\
$I_0{\cal O}_{xy^\prime}$ & $ 2\Im\left(-{\cal M}_2{\cal M}_3^* + 
     {\cal M}_1{\cal M}_4^*\right) 
$ & $ -2\Im\left(b_1b_3^* - b_2b_4^*\right) $& $\left\{x;y^\prime\right\}$ & \\
$I_0{\cal O}_{xz^\prime}$ & $ 2\Re\left({\cal M}_1{\cal M}_3^* -  
     {\cal M}_2{\cal M}_4^*\right) 
$ & $ 2\Im\left(b_2b_3^* + b_1b_4^*\right) $& $\left\{x;z^\prime\right\}$ & \\
$I_0{\cal O}_{yx^\prime}$ & $ 2\Im\left({\cal M}_2{\cal M}_3^*+ 
      {\cal M}_1{\cal M}_4^*\right) 
$ & $ 2\Im\left(b_1b_2^* - b_3b_4^*\right) $ & $\left\{y;x^\prime\right\}$ &\\
$I_0{\cal O}_{yy^\prime}$ & $ 2\Re\left(-{\cal M}_2{\cal M}_3^* +  
     {\cal M}_1{\cal M}_4^*\right) 
$ & $ \left|b_1\right|^2 - \left|b_2\right|^2 - 
     \left|b_3\right|^2 + \left|b_4\right|^2 $& $\left\{y;y^\prime\right\}$ & \\
$I_0{\cal O}_{yz^\prime}$ & $ -2\Im\left({\cal M}_1{\cal M}_3^* - 
      {\cal M}_2{\cal M}_4^*\right) 
$ & $ 2\Re\left(-b_1b_2^* + b_3b_4^*\right) $& $\left\{y;z^\prime\right\}$ & \\
$I_0{\cal O}_{zx^\prime}$ & $ 2\Re\left({\cal M}_1{\cal M}_2^*  - 
     {\cal M}_3{\cal M}_4^*\right) 
$ & $ 2\Im\left(b_2b_3^* - b_1b_4^*\right) $& $\left\{z;x^\prime\right\}$ & \\
$I_0{\cal O}_{zy^\prime}$ & $ -2\Im\left({\cal M}_1{\cal M}_2^* - 
      {\cal M}_3{\cal M}_4^*\right) 
$ & $ 2\Re\left(- b_1b_3^* + b_2b_4^*\right) $ & $\left\{z;y^\prime\right\}$ &\\
$I_0{\cal O}_{zz^\prime}$ & $ -\left|{\cal M}_1\right|^2 + 
     \left|{\cal M}_2\right|^2 + \left|{\cal M}_3\right|^2 - 
     \left|{\cal M}_4\right|^2
$ & $ 2\Re\left(b_2b_3^* + b_1b_4^*\right)$& $\left\{z;z^\prime\right\}$ &\\[+10pt] 
\hline
\end{tabular}
\end{center}
\end{table}

The sixteen polarization observables that are possible are shown in table 
\ref{tab:singlepolobspi}. These sixteen quantities are not all independent, as a number of relationships
can be obtained among them. We first list six relationships that arise from considering the absolute
magnitudes of the transversity amplitudes. These are
\beqy
\left(P_{x^\prime}\pm{\cal O}_{yx^\prime}\right)^2+\left(P_{z^\prime}\pm{\cal O}_{yz^\prime}\right)^2&=&
\left(1\pm P_y\right)^2-\left(P_{y^\prime}\pm{\cal O}_{yy^\prime}\right)^2,\nonumber\\
\left(P_x\pm{\cal O}_{xy^\prime}\right)^2+\left(P_z\pm{\cal O}_{zy^\prime}\right)^2&=&
\left(1\pm P_{y^\prime}\right)^2-\left(P_y\pm{\cal O}_{yy^\prime}\right)^2,\nonumber\\
\left({\cal O}_{xx^\prime}\pm{\cal O}_{zz^\prime}\right)^2+\left({\cal O}_{xz^\prime}\mp{\cal O}_{zx^\prime}\right)^2&=&
\left(1\pm {\cal O}_{yy^\prime}\right)^2-\left(P_y\pm P_{y^\prime}\right)^2.
\eeqy
These six identities may be used to place limits on the absolute magnitudes of some of the
observables. Since the left-hand sides of all six of these equations are positive definite, we obtain the
inequalities
\beqy
\left|1\pm P_y\right|&\ge&\left|P_{y^\prime}\pm{\cal O}_{yy^\prime}\right|,\nonumber\\
\left|1\pm P_{y^\prime}\right|&\ge&\left|P_y\pm{\cal O}_{yy^\prime}\right|,\nonumber\\
\left|1\pm {\cal O}_{yy^\prime}\right|&\ge&\left|P_y\pm P_{y^\prime}\right|.\nonumber
\eeqy
These are similar to the inequalities usually reported in the literature for pion photoproduction, for instance. 
In fact, simple
rearrangement of the equations above allow a larger set of inequalities to be
written. These are 
\beqy
\left|1\pm P_y\right|&\ge&\left\{\left|P_{y^\prime}\pm{\cal O}_{yy^\prime}\right|,
\left|P_{x^\prime}\pm{\cal O}_{yx^\prime}\right|,
\left|P_{z^\prime}\pm{\cal O}_{yz^\prime}\right|\right\},\nonumber\\
\left|1\pm P_{y^\prime}\right|&\ge&\left\{\left|P_y\pm{\cal O}_{yy^\prime}\right|,
\left|P_x\pm{\cal O}_{xy^\prime}\right|,\left|P_z\pm{\cal O}_{zy^\prime}\right|\right\},\nonumber\\
\left|1\pm {\cal O}_{yy^\prime}\right|&\ge&\left\{\left|P_y\pm P_{y^\prime}\right|,
\left|{\cal O}_{xx^\prime}\pm{\cal O}_{zz^\prime}\right|,
\left|{\cal O}_{xz^\prime}\mp{\cal O}_{zx^\prime}\right|\right\}.\nonumber
\eeqy
Further manipulation of these inequalities leads to
\beqy
1+ P_y^2&\ge&\left\{P_{y^\prime}^2+{\cal O}_{yy^\prime}^2,
P_{x^\prime}^2+{\cal O}_{yx^\prime}^2,
P_{z^\prime}^2+{\cal O}_{yz^\prime}^2\right\},\nonumber\\
1+ P_{y^\prime}^2&\ge&\left\{P_y^2+{\cal O}_{yy^\prime}^2,
P_x^2+{\cal O}_{xy^\prime}^2,P_z^2+{\cal O}_{zy^\prime}^2\right\},\nonumber\\
1+ {\cal O}_{yy^\prime}^2&\ge&\left\{P_y^2+ P_{y^\prime}^2,
{\cal O}_{xx^\prime}^2+{\cal O}_{zz^\prime}^2,
{\cal O}_{xz^\prime}^2+{\cal O}_{zx^\prime}^2\right\}.\nonumber
\eeqy

Of the sixteen observables, ten are therefore independent. We can further reduce the number of
independent observables by using the relationships that exist among the phases of the 
transversity amplitudes. Since there will be one overall phase that is unmeasurable, only three of the
relative phases are independent. This means that three relative phases can be eliminated, providing three
more relationships among the observables, leaving seven independent observables. 
The three identities obtained this way
can be displayed in a number of different ways, depending on, for instance, which phases (or phase differences) are chosen to be the independent
ones. Writing $b_i=\rho_ie^{i\phi_i}$, and defining all phase differences relative to $\phi_4$, we obtain
\beqy
-\frac{P_{x^\prime}+{\cal O}_{yx^\prime}}{P_{z^\prime}+{\cal O}_{yz^\prime}}&=&
\frac{\left({\cal O}_{xz^\prime}-{\cal O}_{zx^\prime}\right)\left({\cal O}_{zy^\prime}-P_z\right)-
\left({\cal O}_{xx^\prime}+{\cal O}_{zz^\prime}\right)\left({\cal O}_{xy^\prime}-P_z\right)}
{\left({\cal O}_{xx^\prime}+{\cal O}_{zz^\prime}\right)\left({\cal O}_{zy^\prime}-P_z\right)+
\left({\cal O}_{xz^\prime}-{\cal O}_{zx^\prime}\right)\left({\cal O}_{xy^\prime}-P_x\right)},\nonumber \\
\frac{P_x+{\cal O}_{xy^\prime}}{P_z+{\cal O}_{zz^\prime}}&=&
\frac{\left({\cal O}_{xz^\prime}-{\cal O}_{zx^\prime}\right)\left({\cal O}_{yz^\prime}-P_{z^\prime}\right)-
\left({\cal O}_{xx^\prime}+{\cal O}_{zz^\prime}\right)\left(P_{x^\prime}-{\cal O}_{yx^\prime}\right)}
{\left({\cal O}_{xx^\prime}+{\cal O}_{zz^\prime}\right)\left({\cal O}_{yz^\prime}-P_{z^\prime}\right)+
\left({\cal O}_{xz^\prime}-{\cal O}_{zx^\prime}\right)\left(P_{x^\prime}-{\cal O}_{yx^\prime}\right)},\nonumber \\
\frac{{\cal O}_{xz^\prime}+{\cal O}_{zx^\prime}}{{\cal O}_{zz^\prime}-{\cal O}_{xx^\prime}}&=&
\frac{\left({\cal O}_{xy^\prime}-P_x\right)\left({\cal O}_{yz^\prime}-P_{z^\prime}\right)-
\left(P_{x^\prime}-{\cal O}_{yx^\prime}\right)\left({\cal O}_{zy^\prime}-P_z\right)}
{\left({\cal O}_{zy^\prime}-P_z\right)\left({\cal O}_{yz^\prime}-P_{z^\prime}\right)+
\left({\cal O}_{xy^\prime}-P_x\right)\left(P_{x^\prime}-{\cal O}_{yx^\prime}\right)}.
\eeqy
We emphasize here that
we have only considered the relationships among the observables. The number seven does not necessarily
represent a `minimal set' that must be measured for the so-called `complete' experiment. We postpone 
such a discussion until later in this section.

It is interesting to note that we can obtain a different set of relationships among the
observables by consideration of the helicity amplitudes instead of the transversity ones.
Proceeding in this way, the relationships obtained are
\beqy
\left(P_x\pm{\cal O}_{xz^\prime}\right)^2+\left(P_y\pm{\cal O}_{yz^\prime}\right)^2&=&
\left(1\pm P_{z^\prime}\right)^2-\left(P_z\pm{\cal O}_{zz^\prime}\right)^2,\nonumber\\
\left({\cal O}_{xx^\prime}\pm{\cal O}_{yy^\prime}\right)^2+
\left({\cal O}_{xy^\prime}\mp{\cal O}_{yx^\prime}\right)^2&=&
\left(1\pm {\cal O}_{zz^\prime}\right)^2-\left(P_z\pm P_{z^\prime}\right)^2,\nonumber\\
\left(P_{x^\prime}\pm{\cal O}_{zx^\prime}\right)^2+\left(P_{y^\prime}\pm{\cal O}_{zy^\prime}\right)^2&=&
\left(1\pm P_z\right)^2-\left(P_{z^\prime}\pm {\cal O}_{zz^\prime}\right)^2.
\eeqy
The corresponding inequalities obtained from these are
\beqy
\left|1\pm P_{z^\prime}\right|&\ge&\left\{\left|P_z\pm{\cal O}_{zz^\prime}\right|,
\left|P_x\pm{\cal O}_{xz^\prime}\right|,\left|P_y\pm{\cal O}_{yz^\prime}\right|\right\},\nonumber\\
\left|1\pm {\cal O}_{zz^\prime}\right|&\ge&\left\{\left|P_z\pm P_{z^\prime}\right|,
\left|{\cal O}_{xx^\prime}\pm{\cal O}_{yy^\prime}\right|,
\left|{\cal O}_{xy^\prime}\mp{\cal O}_{yx^\prime}\right|\right\},\nonumber\\
\left|1\pm P_z\right|&\ge&\left\{\left|P_{z^\prime}\pm {\cal O}_{zz^\prime}\right|,
\left|P_{x^\prime}\pm{\cal O}_{zx^\prime}\right|,\left|P_{y^\prime}\pm{\cal
O}_{zy^\prime}\right|\right\}
\eeqy
and
\beqy
1+ P_{z^\prime}^2&\ge&\left\{P_z^2+{\cal O}_{zz^\prime}^2,
P_x^2+{\cal O}_{xz^\prime}^2,P_y^2+{\cal O}_{yz^\prime}^2\right\},\nonumber\\
1+ {\cal O}_{zz^\prime}^2&\ge&\left\{P_z^2+ P_{z^\prime}^2,
{\cal O}_{xx^\prime}^2+{\cal O}_{yy^\prime}^2,
{\cal O}_{xy^\prime}^2+{\cal O}_{yx^\prime}^2\right\},\nonumber\\
1+ P_z^2&\ge&\left\{P_{z^\prime}^2+ {\cal O}_{zz^\prime}^2,
P_{x^\prime}^2+{\cal O}_{zx^\prime}^2,P_{y^\prime}^2+{\cal
O}_{zy^\prime}^2\right\}.
\eeqy
In a similar manner, a set of relationships may be obtained by considering the
phases of the helicity amplitudes.

\subsection{Required Experimental Measurements in $\pi N\to\pi\pi N$}

Information on baryon spectroscopy is obtained from processes like $\pi N\to\pi N$ by extracting the
helicity or transversity (or partial wave) amplitudes for the process. These amplitudes are then 
interpreted in terms of baryon resonances. There is therefore a great deal of interest in knowing how
many measurements must  be made at each kinematic point, in order to extract the amplitudes. For this
discussion, we focus on the process $\pi N\to\pi\pi N$, since such discussions have already been
documented for $\pi N\to\pi N$.

If we write $b_i=\rho_ie^{i\phi_i}$, then the quantities $I_0$, $P_y$, $P_{y^\prime}$ and 
${\cal O}_{yy^\prime}$ must be measured at each kinematic point to provide the information needed to
extract the $\rho_i$ unambiguously. In the bilinear combinations of transversity amplitudes, there are
six phase differences that occur, but only three of these are independent. Any three can be chosen, so
we discuss $\phi_{12}\equiv\phi_1-\phi_2$, $\phi_{34}\equiv\phi_3-\phi_4$ and
$\phi_{23}\equiv\phi_2-\phi_3$.

To access $\phi_{12}$, two of the four quantities $P_{x^\prime}$, $P_{z^\prime}$, ${\cal O}_{yx^\prime}$
and ${\cal O}_{yz^\prime}$ should be measured at each kinematic point. The pair of measurements $P_{x^\prime}$
and ${\cal O}_{yx^\prime}$, or $P_{z^\prime}$ and ${\cal O}_{yz^\prime}$, would provide `cleaner'
solutions for $\phi_{12}$. Note that these measurement would also provide $\phi_{34}$, and both of these
phase differences would be subject to the well-known `quadrant ambiguities' \cite{bds}. 

This leaves one more phase difference to be determined. If we choose this to be $\phi_{23}$, then
measurement of one of ${\cal O}_{xx^\prime}$, ${\cal O}_{zx^\prime}$, ${\cal O}_{xz^\prime}$ or
${\cal O}_{zz^\prime}$ will allow its extraction. In order to do this, however, the other phase that
occurs in these observables, $\phi_{14}$, will have to be written in terms of the two phases already
extracted and $\phi_{23}$ as
\begin{equation}
\phi_{14}=\phi_1-\phi_4=\phi_1-\phi_2+\phi_2-\phi_3+\phi_3-\phi_4=\phi_{12}+\phi_{23}+\phi_{34}.
\end{equation}
Then, the only unknown in the measured quantity would be $\phi_{23}$.

A similar analysis can be made in terms of the helicity amplitudes. In this case, $I_0$, $P_z$, 
$P_{z^\prime}$ and 
${\cal O}_{zz^\prime}$ must be measured at each kinematic point in order to determine the magnitudes of
the helicity amplitudes. Two measurements from among $P_{x^\prime}$, $P_{y^\prime}$, 
${\cal O}_{zx^\prime}$
and ${\cal O}_{zy^\prime}$ will provide enough information to determine two of the relative phases, for
instance, and one measurement from among $P_{x}$, $P_{y}$, ${\cal O}_{xz^\prime}$
and ${\cal O}_{yz^\prime}$ will provide enough information to determine the last phase needed.

The bottom line is that in order to extract reliable information on baryon properties, the helicity or 
transversity amplitudes must be extracted with some degree of certainty, and this can only be done if
at least seven judiciously chosen measurements are performed at each kinematic point. This 
also means that both single and double polarization measurements will be essential. This is
similar to the conclusion of ref. \cite{bds} in their analysis of pion photoproduction, and is
independent of whether the observables are described in terms of helicity, transversity, or
other amplitudes.

\subsection{Parity Conservation}

The properties of the helicity and transversity amplitudes for a process $a+b\to
c+d$ are well known. For $\pi N\to\pi N$, the relationships among the helicity
amplitudes are written \cite{jacobwick}
\begin{equation}
{\cal M}_{-\lambda_N-\lambda_N^\prime}(\theta)=(-1)^{\lambda_N-\lambda_N^\prime}
{\cal M}_{\lambda_N\lambda_N^\prime}(\theta),
\end{equation}
where $\theta$ is as defined in eqn. (\ref{kinepnpn}). The corresponding relationships for 
transversity amplitudes are \cite{kotanski}
\beq
b_{\tau_N\tau_N^\prime}(\theta)=(-1)^{\tau_N-\tau_N^\prime}
b_{\tau_N\tau_N^\prime}(\theta).
\eeq
Parity conservation therefore means that some of the transversity amplitudes vanish
exactly.

In general, a minimum of three angles are needed to describe the scattering amplitude
for a process $a+b\to c+d+e$. For the specific case of $\pi N\to\pi\pi N$, we 
choose these angles to be as defined in eqns. (\ref{kinepnppn1}) and
(\ref{kinepnppn2}). The relationships that arise among the helicity amplitudes may then be written
\begin{equation}
{\cal M}_{-\lambda_N-\lambda_N^\prime}(\theta,\theta_1,\phi_1)=(-1)^{\lambda_N-\lambda_N^\prime}
{\cal M}_{\lambda_N\lambda_N^\prime}(\theta,\theta_1,2\pi-\phi_1).
\end{equation}
These relations can not be used to decrease the number of independent helicity
amplitudes, but they can be used to determine which of the observables are even or odd under
the transformation $\phi_1\leftrightarrow 2\pi-\phi_1$. 

\subsection{Construction of Transition Amplitudes}

\subsubsection{$\pi N\to\pi N$}

For this process, ${\cal A}$ must be a scalar, and $\vec{\cal B}$ an axial 
vector. With the kinematics for this process as previously defined, we can 
write ${\cal A}$ and $\vec{\cal B}$ as
\beqy
{\cal A}&=& \alpha,\nonumber\\
\vec{\cal B}&=& \beta \frac{\hat{k}\times\hat{q}}{\left|\hat{k}\times\hat{q}\right|},
\eeqy
where $\alpha$ and $\beta$ are scalar quantities that contain all of the details of whatever 
model is constructed to describe the process. These can be compared to the form
usually written for this process \cite{piona}, namely
\beq
i{\cal M}=\chi^\dag\left(f+g\vec{\sigma}\cdot\hat{n}\right),
\eeq
where $\hat{n}=\frac{\hat{k}\times\hat{q}}{\left|\hat{k}\times\hat{q}\right|}$.
This means that we can identify $\alpha=f$ and $\beta=g$.
With these kinematics, ${\cal B}_1={\cal B}_3=0$, leading to 
\beq
{\cal M}_1={\cal M}_4,\,\,\,\, {\cal M}_2=-{\cal M}_3
\eeq
in the helicity basis, or
\beq
b_2=b_3=0
\eeq 
in the transversity basis. Two of the transversity
amplitudes (the transversity-`flip' amplitudes) vanish identically (as expected), 
meaning that this process is exactly `transversity conserving'.
The relationships among the helicity amplitudes expected from considerations of parity
symmetry are therefore obtained. Many of the polarization observables now become 
redundant, or vanish identically:
\beqy
&&P_x=P_z=P_{x^\prime}=P_{z^\prime}={\cal O}_{xy^\prime}={\cal O}_{yx^\prime}=
{\cal O}_{yz^\prime}={\cal O}_{zy^\prime}=0,\nonumber\\
&&I_0=\left|{\cal M}_1\right|^2+ \left|{\cal M}_2\right|^2=\left|b_1\right|^2+
\left|b_2\right|^2=
{\cal O}_{yy^\prime},\nonumber\\
&&P_y=2\Im\left({\cal M}_1{\cal M}_2^*\right)=\left|b_1\right|^2-
\left|b_2\right|^2=P_{y^\prime},\nonumber\\
&&{\cal O}_{xx^\prime}=-\left|{\cal M}_1\right|^2 + 
   \left|{\cal M}_2\right|^2=2\Re\left(b_1b_4^*\right)
{\cal O}_{zz^\prime},\nonumber\\
&&{\cal O}_{xz^\prime}=
-2\Re\left({\cal M}_1{\cal M}_2^*\right)=2\Im\left(b_1b_4^*\right)-{\cal O}_{zx^\prime},
\eeqy
and there are only four independent observables, as expected. From consideration of the
transversity amplitudes, it is `obvious' why ${\cal O}_{yy^\prime}$ and $I_0$ are equal.
The relationships among observables reduces to a single relationship in this case, namely
\beq
P_y^2+{\cal O}_{xx^\prime}^2+{\cal O}_{xz^\prime}^2=1.
\eeq

We note that the convention has been to choose
both sets of axes for this process to be the same. This introduces 
explicit factors of $\cos{\theta}$ and $\sin{\theta}$ into the observables. 
If we choose unrotated primed axes, the relationships among the observables we
have defined, and those found in the literature, are,
\beqy
R&=&{\cal O}_{xx^\prime}\cos{\theta}-{\cal O}_{xz^\prime}\sin{\theta},\nonumber\\
A&=&{\cal O}_{xx^\prime}\sin{\theta}+{\cal O}_{xz^\prime}\cos{\theta}.\nonumber
\eeqy
In terms of $A$ and $R$, the identity that must be satisfied is
\beq
P_y^2+R^2+A^2=1.
\eeq

\subsubsection{$\pi N\to\pi\pi N$}

For this process, there are three independent three-momenta, which we can choose to be $\vec{k}$,
$\vec{p}_2$ and $\vec{q}_1$, where $\vec{q}_1$ is the momentum of one of the 
final pions. These have all been defined previously.
In this case, ${\cal A}$ must be a pseudoscalar quantity, while ${\cal B}$ must be a vector. The
only possibilities are
\beqy
{\cal A}&=& \alpha \hat{k}\cdot \hat{p}_2\times\hat{q}_1,\nonumber\\
\vec{\cal B}&=& \beta_1 \hat{k}+\beta_2\hat{p}_2+\beta_3\hat{q}_1,
\eeqy
where $\alpha$ and the $\beta_i$ depend only on scalar products of the vectors in the problem. For this
case, $\vec{\cal B}$ has $x$, $y$ and $z$ components, and none of the four helicity amplitudes
are independent. Furthermore, none of the polarization observables vanish exactly, and all are
independent. However, using the properties of the helicity amplitudes, we can predict
that the observables $P_x,\,\,P_z,\,\,P_{x^\prime},\,\,P_{z^\prime},\,\,
{\cal O}_{xy^\prime},\,\,{\cal O}_{yx^\prime},\,\,
{\cal O}_{yz^\prime}$ and ${\cal O}_{zy^\prime}$ are all odd under the transformation
$\phi_1\leftrightarrow 2\pi-\phi_1$. The other eight observables are all even
under this transformation.

\section{Observables in $\gamma N \to \pi N$ and $\gamma N \to\pi\pi N$}

We can treat these two processes in a framework similar to that used for $\pi N\to\pi N$
and $\pi N\to\pi\pi N$ by writing
\beq
i{\cal M}=\chi^\dag\left({\cal A}_j+ \sigma_i{\cal B}_{ij}\right)\phi\varepsilon_j,
\eeq
where $\vec{\varepsilon}$ is the polarization vector of the initial photon, ${\cal A}_j$ are
the components of a vector ($\gamma N\to\pi\pi N$) or an axial vector ($\gamma N\to\pi N$), and
${\cal B}_{ij}$ is a tensor ($\gamma N\to\pi N$) or pseudotensor ($\gamma N\to\pi\pi N$). The
amplitude takes the form
\beq
i{\cal M}=\chi^\dag\left(\begin{array}{cc} A_{+j}& B_{-j}\\
B_{+j}& A_{-j}\end{array}\right)\phi\varepsilon_j,
\eeq
where
\beq
A_{\pm j}\equiv {\cal A}_j\pm {\cal B}_{3j}, \,\,\,\, B_{\pm j}\equiv {\cal
B}_{1j}\pm i{\cal B}_{2j},
\eeq
in analogy with our treatment of $\pi N\to\pi N$ and $\pi N\to\pi\pi N$.

Defining the helicity amplitudes
\beqy
i{\cal M}_{++}^\lambda&\equiv {\cal M}_1^\lambda,\nonumber\\
i{\cal M}_{+-}^\lambda&\equiv {\cal M}_2^\lambda,\nonumber\\
i{\cal M}_{-+}^\lambda&\equiv {\cal M}_3^\lambda,\nonumber\\
i{\cal M}_{--}^\lambda&\equiv {\cal M}_4^\lambda,
\eeqy
where $\lambda$ is the helicity of the photon, and the transversity amplitudes
\beqy
ib_{++}^\lambda&\equiv& b_1^\lambda,\nonumber\\
ib_{+-}^\lambda&\equiv& b_2^\lambda,\nonumber\\
ib_{-+}^\lambda&\equiv& b_3^\lambda,\nonumber\\
ib_{--}^\lambda&\equiv& b_4^\lambda,
\eeqy
we find
\beqy \label{hell0}
{\cal A}_j\varepsilon_j(\lambda)&=&\frac{1}{2}\left[\left({\cal M}_1^\lambda+{\cal M}_4^\lambda\right)\cos{\frac{\theta}{2}}+
\left({\cal M}_2^\lambda-{\cal M}_3^\lambda\right)\sin{\frac{\theta}{2}}\right]=
\frac{1}{2}\left(b_1^\lambda+b_4^\lambda\right),\nonumber\\
{\cal B}_{1j}\varepsilon_j(\lambda)&=&\frac{1}{2}\left[\left({\cal M}_2^\lambda+{\cal M}_3^\lambda\right)\cos{\frac{\theta}{2}}+
\left({\cal M}_4^\lambda-{\cal M}_1^\lambda\right)\sin{\frac{\theta}{2}}\right]=
\frac{i}{2}\left(b_2^\lambda-b_3^\lambda\right),\nonumber\\
{\cal B}_{2j}\varepsilon_j(\lambda)&=&\frac{i}{2}\left[\left({\cal M}_2^\lambda-{\cal M}_3^\lambda\right)\cos{\frac{\theta}{2}}-
\left({\cal M}_1^\lambda+{\cal M}_4^\lambda\right)\sin{\frac{\theta}{2}}\right]=
\frac{1}{2}\left(b_1^\lambda-b_4^\lambda\right),\nonumber\\
{\cal B}_{3j}\varepsilon_j(\lambda)&=&\frac{1}{2}\left[\left({\cal
M}_4^\lambda-{\cal M}_1^\lambda\right)\cos{\frac{\theta}{2}}-
\left({\cal M}_2^\lambda+{\cal M}_3^\lambda\right)\sin{\frac{\theta}{2}}\right]=
\frac{1}{2}\left(b_2^\lambda+b_3^\lambda\right).
\eeqy
Note that the amplitudes $b_i^\lambda$ are not strictly transversity amplitudes,
as the photon spin is still quantized along its direction of motion. Quantizing along the
transverse direction requires construction of the combinations ${\cal D}_i^\pm=
b_i^+\pm b_i^-$.

The transversity spinors of eqn. (\ref{transverse}) can be written as linear 
superpositions of the helicity spinors of eqns. (\ref{eq:helicityspin1/2}) and
(\ref{helicity}). The expressions are
\begin{eqnarray} 
\phi^+_T&=&\frac{1}{\sqrt{2}}\left(\phi^+-i\phi^-\right)
=\frac{1}{\sqrt{2}}e^{i\theta/2}\left(\chi^+-i\chi^-\right),\nonumber\\
\phi^-_T&=&\frac{1}{\sqrt{2}}\left(\phi^++i\phi^-\right)
=\frac{1}{\sqrt{2}}e^{-i\theta/2}\left(\chi^++i\chi^-\right).
\end{eqnarray}
This allows yet another set of amplitudes, $i{\cal
T}_{\tau_N\tau_N^\prime}^{\lambda_\gamma}$, to be defined in terms of the helicity amplitudes. 
These are
\begin{eqnarray}
i{\cal T}_{++}^\pm&\equiv& {\cal T}_1^\pm=\frac{1}{2}e^{-i\theta/2}\left[{\cal M}_1^\pm+
{\cal M}_4^\pm+i\left({\cal M}_2^\pm-{\cal M}_3^\pm\right)\right],\nonumber\\
i{\cal T}_{+-}^\pm&\equiv& {\cal T}_2^\pm=\frac{1}{2}e^{i\theta/2}\left[{\cal M}_1^\pm-
{\cal M}_4^\pm-i\left({\cal M}_2^\pm+{\cal M}_3^\pm\right)\right],\nonumber\\
i{\cal T}_{-+}^\pm&\equiv& {\cal T}_3^\pm=\frac{1}{2}e^{-i\theta/2}\left[{\cal M}_1^\pm-
{\cal M}_4^\pm+i\left({\cal M}_2^\pm+{\cal M}_3^\pm\right)\right],\nonumber\\
i{\cal T}_{--}^\pm&\equiv& {\cal T}_4^\pm=\frac{1}{2}e^{i\theta/2}\left[{\cal M}_1^\pm+
{\cal M}_4^\pm-i\left({\cal M}_2^\pm-{\cal M}_3^\pm\right)\right].
\end{eqnarray}
Full transversity amplitudes can be constructed from these as ${\cal D}_i^{\tau_\gamma}=
{\cal T}_i^+\pm{\cal T}_i^-$. For $\gamma N\to\pi N$, the resulting amplitudes are similar 
to those found in the literature, but the phases $e^{\pm
i\theta/2}$ are absent from the published forms (see ref. \cite{storrow}, page 270).

We can write the reaction rate $I$, as
\beqy
\rho_fI&=&I_0\left\{\left(1+\vec{\Lambda}_i\cdot\vec{P}+\vec{\sigma}\cdot\vec{P}^\prime+
\Lambda_i^\alpha\sigma^{\beta^\prime}{\cal O}_{\alpha\beta^\prime}\right)\right.\nonumber\\
&&+\left.\delta_\odot\left(I^\odot+\vec{\Lambda}_i\cdot\vec{P}^\odot+\vec{\sigma}\cdot\vec{P}^{\odot\prime}+
\Lambda_i^\alpha\sigma^{\beta^\prime}{\cal O}^\odot_{\alpha\beta^\prime}\right)\right.\nonumber\\
&&+\left.\delta_\ell\left[\sin{2\beta}\left(I^s+\vec{\Lambda}_i\cdot\vec{P}^s+\vec{\sigma}\cdot\vec{P}^{s\prime}+
\Lambda_i^\alpha\sigma^{\beta^\prime}{\cal O}^s_{\alpha\beta^\prime}\right)\right.\right.\nonumber\\
&&+\left.\left.\cos{2\beta}\left(I^c+\vec{\Lambda}_i\cdot\vec{P}^c+\vec{\sigma}\cdot\vec{P}^{c\prime}+
\Lambda_i^\alpha\sigma^{\beta^\prime}{\cal O}^c_{\alpha\beta^\prime}\right)\right]\right\}\nonumber\\
\eeqy
where $\vec{P}$ represents the polarization asymmetry that arises if the target nucleon is
polarized, $\rho_f=\frac{1}{2}\left(1+\vec{\sigma}\cdot \vec{P}^\prime\right)$ is the density 
matrix of the recoiling nucleon, and ${\cal
O}_{\alpha\beta^\prime}$ is the observable if both the target and recoil polarization are measured.
The primes indicate that the recoil observables are measured with respect to a set of axes
$x^\prime,\,y^\prime,\,z^\prime$, in which $z^\prime$ is along the direction of motion of the
recoiling nucleon, and $y^\prime=y$. $\delta_\odot$ is the degree of circular polarization in the photon beam, while 
$\delta_\ell$ is the degree of linear polarization, with the direction of
polarization being at an angle $\beta$ to the $x$-axis.

The polarization observables that arise for these two processes are given in the four tables
that follow. Note that there are 64 polarization observables in general.

\squeezetable

\begin{table}[p]
\caption{Polarization observables of single and double pion photoproduction
in terms of the helicity and transversity amplitudes. These observables
arise with an unpolarized photon beam. Variables labeled with a T require a
polarized target with recoil polarization unmeasured, while those labeled with
an R require an unpolarized target, but the recoil polarization is measured.
Those denoted TR require polarized targets, with recoil polarization measured.
The measurements required are shown by the set $\left\{b,t,r\right\}$, which 
denote the component of the target (t)
or  recoil (r) polarization that must known or measured. For the target
polarization, the $x$-, $y$- and $z$-axes are as defined in the text. The 
$x^\prime$-, $y^\prime$- and $z^\prime$-axes are also defined in the text, as is
the notation for the transversity amplitudes.\label{tab:doublepolobs}}
\begin{center}
\begin{tabular}{c|l|l|c|c} 
\hline Observable & Helicity Form  & Transversity Form & Expt. & Type \\ 
\hline 
$I_0$ & $\begin{array}{l}\left|{\cal M}_1^-\right|^2 + \left|{\cal 
M}_1^+\right|^2 + \left|{\cal M}_2^-\right|^2 + \left|{\cal 
M}_2^+\right|^2 \\+ \left|{\cal M}_3^-\right|^2 + \left|{\cal 
M}_3^+\right|^2 + \left|{\cal M}_4^-\right|^2 + \left|{\cal 
M}_4^+\right|^2\end{array}$ & 
 $ \begin{array}{l}\left|b_1^-\right|^2 + \left|b_1^+\right|^2 + \left|b_2^-
\right|^2 + \left|b_2^+\right|^2 \\+ \left|b_3^-\right|^2 + \left|
b_3^+\right|^2 + \left|b_4^-\right|^2 + \left|b_4^+\right|^2\end{array}$ & $\left\{-;-;-\right\}$ & \\ [+9pt]
$I_0P_x$ & $ 2\Re\left({\cal M}_1^-{\cal M}_3^{-*} + 
{\cal M}_1^+{\cal M}_3^{+*} + 
{\cal M}_2^-{\cal M}_4^{-*} +{\cal M}_2^+{\cal M}_4^{+*}\right)$ & 
 $ -2\Im\left(b_1^-b_3^{-*} + b_1^+b_3^{+*}+ 
b_2^-b_4^{-*} + b_2^+b_4^{+*}\right)$ & $\left\{-;x;-\right\}$ & T \\ [+9pt]
$I_0P_y$ & $ -2\Im\left({\cal M}_1^-{\cal M}_3^{-*} + 
{\cal M}_1^+{\cal M}_3^{+*} + 
{\cal M}_2^-{\cal M}_4^{-*} +{\cal M}_2^+{\cal M}_4^{+*}
\right)$ & 
 $ \begin{array}{l}\left|b_1^-\right|^2 + \left|b_1^+\right|^2 + \left|b_2^-
\right|^2 + \left|b_2^+\right|^2 \\- \left|b_3^-\right|^2 - \left|b_3^+
\right|^2 - \left|b_4^-\right|^2 - \left|b_4^+\right|^2\end{array}$ & $\left\{-;y;-\right\}$ & \\ [+9pt]
$I_0P_z$ & $\begin{array}{l}-\left|{\cal M}_1^-\right|^2 - \left|{\cal 
M}_1^+\right|^2 - \left|{\cal M}_2^-\right|^2 - \left|{\cal 
M}_2^+\right|^2 \\ + \left|{\cal M}_3^-\right|^2 + \left|{\cal 
M}_3^+\right|^2 + \left|{\cal M}_4^-\right|^2 + \left|{\cal 
M}_4^+\right|^2\end{array}$ & 
 $ -2\Re\left(b_1^-b_3^{-*} + b_1^+b_3^{+*} +
b_2^-b_4^{-*} + b_2^+b_4^{+*}\right)$ & $\left\{-;z;-\right\}$ & \\ [+9pt]
$I_0P_{x^\prime}$ & $ -2\Re\left({\cal M}_1^-{\cal M}_2^{-*} +
{\cal M}_1^+{\cal M}_2^{+*} +
{\cal M}_3^-{\cal M}_4^{-*} +{\cal M}_3^+{\cal M}_4^{+*}\right)$ & 
 $ 2\Im\left( b_1^-b_2^{-*} + b_1^+b_2^{+*} + 
b_3^-b_4^{-*} + b_3^+b_4^{+*}\right)$ & $\left\{-;-;x^\prime\right\}$ & R \\ [+9pt]
$I_0P_{y^\prime}$ & $ 2\Im\left({\cal M}_1^-{\cal M}_2^{-*} +
{\cal M}_1^+{\cal M}_2^{+*} +
{\cal M}_3^-{\cal M}_4^{-*} +{\cal M}_3^+{\cal M}_4^{+*}\right)$ & 
 $ \begin{array}{l}\left|b_1^-\right|^2 + \left|b_1^+\right|^2
 - \left|b_2^-\right|^2 - \left|b_2^+\right|^2 \\+ \left|b_3^-\right|^2 + 
 \left|b_3^+\right|^2 - \left|b_4^-\right|^2 - \left|b_4^+\right|^2\end{array}$ & $\left\{-;-;y^\prime\right\}$ & \\ [+9pt]
$I_0P_{z^\prime}$ & $ \begin{array}{l}\left|{\cal M}_1^-\right|^2 + \left|{\cal 
M}_1^+\right|^2 - \left|{\cal M}_2^-\right|^2 - \left|{\cal 
M}_2^+\right|^2 \\+ \left|{\cal M}_3^-\right|^2 + \left|{\cal 
M}_3^+\right|^2 - \left|{\cal M}_4^-\right|^2 - \left|{\cal 
M}_4^+\right|^2\end{array}$ & 
 $ -2\Re\left(b_1^-b_2^{-*} + b_1^+b_2^{+*} +  
b_3^-b_4^{-*} + b_3^+b_4^{+*}\right)$ & $\left\{-;-;z^\prime\right\}$ & \\ [+9pt]
$I_0{\cal O}_{xx^\prime}$ & $ -2\Re\left({\cal M}_2^-{\cal M}_3^{-*} + 
{\cal M}_2^+{\cal M}_3^{+*} +
{\cal M}_1^-{\cal M}_4^{-*} +{\cal M}_1^+{\cal M}_4^{+*}\right)$ & 
 $ 2\Re\left(-b_2^-b_3^{-*} - b_2^+b_3^{+*} + 
b_1^-b_4^{-*} + b_1^+b_4^{+*}\right)$ & $\left\{-;x;x^\prime\right\}$ & TR \\ [+9pt]
$I_0{\cal O}_{xy^\prime}$ & $ 2\Im\left(-{\cal M}_2^-{\cal M}_3^{-*} - 
{\cal M}_2^+{\cal M}_3^{+*} + 
{\cal M}_1^-{\cal M}_4^{-*} +{\cal M}_1^+{\cal M}_4^{+*}\right)$ & 
 $ -2\Im\left(b_1^-b_3^{-*} + b_1^+b_3^{+*}  - 
b_2^-b_4^{-*} - b_2^+b_4^{+*}\right)$ & $\left\{-;x;y^\prime\right\}$ & \\ [+9pt]
$I_0{\cal O}_{xz^\prime}$ & $ 2\Re\left({\cal M}_1^-{\cal M}_3^{-*} + 
{\cal M}_1^+{\cal M}_3^{+*} - 
{\cal M}_2^-{\cal M}_4^{-*} -{\cal M}_2^+{\cal M}_4^{+*}\right)$ & 
 $ 2\Im\left(b_2^-b_3^{-*} + b_2^+b_3^{+*} + 
b_1^-b_4^{-*} + b_1^+b_4^{+*}\right)$ & $\left\{-;x;z^\prime\right\}$ & \\ [+9pt]
$I_0{\cal O}_{yx^\prime}$ & $ 2\Im\left({\cal M}_2^-{\cal M}_3^{-*} + 
{\cal M}_2^+{\cal M}_3^{+*} +
{\cal M}_1^-{\cal M}_4^{-*} +{\cal M}_1^+{\cal M}_4^{+*}\right)$ & 
 $ 2\Im\left(b_1^-b_2^{-*} + b_1^+b_2^{+*} - 
b_3^-b_4^{-*} - b_3^+b_4^{+*}\right)$ & $\left\{-;y;x^\prime\right\}$ & \\ [+9pt]
$I_0{\cal O}_{yy^\prime}$ & $ 2\Re\left(-{\cal M}_2^-{\cal M}_3^{-*} - 
{\cal M}_2^+{\cal M}_3^{+*} + 
{\cal M}_1^-{\cal M}_4^{-*} +{\cal M}_1^+{\cal M}_4^{+*}\right)$ & 
 $ \begin{array}{l}\left|b_1^-\right|^2 + \left|b_1^+\right|^2
 - \left|b_2^-\right|^2 - \left|b_2^+\right|^2 \\- \left|b_3^-\right|^2 
 - \left|b_3^+\right|^2 + \left|b_4^-\right|^2 + \left|b_4^+\right|^2\end{array}$ & $\left\{-;y;y^\prime\right\}$ & \\ [+9pt]
$I_0{\cal O}_{yz^\prime}$ & $ -2\Im\left({\cal M}_1^-{\cal M}_3^{-*} + 
{\cal M}_1^+{\cal M}_3^{+*} - 
{\cal M}_2^-{\cal M}_4^{-*} -{\cal M}_2^+{\cal M}_4^{+*}\right)$ & 
 $ 2\Re\left(-b_1^-b_2^{-*} - b_1^+b_2^{+*} +  
b_3^-b_4^{-*} + b_3^+b_4^{+*}\right)$ &  $\left\{-;y;z^\prime\right\}$ \\ [+9pt]
$I_0{\cal O}_{zx^\prime}$ & $ 2\Re\left({\cal M}_1^-{\cal M}_2^{-*} + 
{\cal M}_1^+{\cal M}_2^{+*} - 
{\cal M}_3^-{\cal M}_4^{-*} -{\cal M}_3^+{\cal M}_4^{+*}\right)$ & 
 $ 2\Im\left(b_2^-b_3^{-*} + b_2^+b_3^{+*} - 
b_1^-b_4^{-*} - b_1^+b_4^{+*}\right)$ & $\left\{-;z;x^\prime\right\}$ & \\ [+9pt]
$I_0{\cal O}_{zy^\prime}$ & $ -2\Im\left({\cal M}_1^-{\cal M}_2^{-*} + 
{\cal M}_1^+{\cal M}_2^{+*} - 
{\cal M}_3^-{\cal M}_4^{-*} -{\cal M}_3^+{\cal M}_4^{+*}\right)$ & 
 $ 2\Re\left(-b_1^-b_3^{-*} - b_1^+b_3^{+*} + 
b_2^-b_4^{-*} + b_2^+b_4^{+*}\right)$ & $\left\{-;z;y^\prime\right\}$ & \\ [+9pt]
$I_0{\cal O}_{zz^\prime}$ & $\begin{array}{l}-\left|{\cal M}_1^-\right|^2 - \left|{
\cal M}_1^+\right|^2 + \left|{\cal M}_2^-\right|^2 + \left|{\cal 
M}_2^+\right|^2 \\ + \left|{\cal M}_3^-\right|^2 + \left|{\cal M}_3^+\right|^2 - 
\left|{\cal M}_4^-\right|^2 - \left|{\cal M}_4^+\right|^2\end{array}$ & 
 $ 2\Re\left(b_2^-b_3^{-*} + b_2^+b_3^{+*} + 
b_1^-b_4^{-*} + b_1^+b_4^{+*}\right)$ & $\left\{-;z;z^\prime\right\}$ & \\ \hline
\hline 
\end{tabular}
\end{center} 
\end{table}

\begin{table}[p]
\caption{Polarization observables of single and double pion photoproduction in 
terms of the helicity and transversity amplitudes. These observables
arise with circularly polarized photons. The notation is as in table 
\ref{tab:doublepolobs}.\label{tab:doublepolobso}}
\begin{center}
\begin{tabular}{c|l|l|c|c} 
\hline Observable & Helicity form & Transversity Form & Expt. & Type\\ 
\hline 
$I_0I^\odot$ & $\begin{array}{l} -\left|{\cal M}_1^-\right|^2 + \left|{\cal 
M}_1^+\right|^2 - \left|{\cal M}_2^-\right|^2 + \left|{\cal 
M}_2^+\right|^2 \\- \left|{\cal M}_3^-\right|^2 + \left|{\cal 
M}_3^+\right|^2 - \left|{\cal M}_4^-\right|^2 + \left|{\cal 
M}_4^+\right|^2\end{array}$ & 
 $\begin{array}{l} -\left|b_1^-\right|^2 + \left|b_1^+\right|^2 - \left|b_2^-
\right|^2 + \left|b_2^+\right|^2 \\- \left|b_3^-\right|^2 + \left|b_3^+
\right|^2 - \left|b_4^-\right|^2 + \left|b_4^+\right|^2\end{array}$ &  $\left\{c;-;-\right\}$ & B$_\odot$\\ [+9pt]
$I_0P_x^\odot$ & $ 2\Re\left(-{\cal M}_1^-{\cal M}_3^{-*} + 
{\cal M}_1^+{\cal M}_3^{+*} - 
{\cal M}_2^-{\cal M}_4^{-*} +{\cal M}_2^+{\cal M}_4^{+*}\right)$ & 
$ 2\Im\left(b_1^-b_3^{-*} - b_1^+b_3^{+*} + 
b_2^-b_4^{-*} - b_2^+b_4^{+*}\right)$ & $\left\{c;x;-\right\}$ & B$_\odot$T \\ [+9pt]
$I_0P_y^\odot$ & $ 2\Im\left({\cal M}_1^-{\cal M}_3^{-*} - 
{\cal M}_1^+{\cal M}_3^{+*} + 
{\cal M}_2^-{\cal M}_4^{-*} -{\cal M}_2^+{\cal M}_4^{+*}\right)$ & 
$ \begin{array}{l}-\left|b_1^-\right|^2 + \left|b_1^+\right|^2 - 
\left|b_2^-\right|^2 + \left|b_2^+\right|^2 \\+ \left|b_3^-\right|^2 - 
\left|b_3^+\right|^2 + \left|b_4^-\right|^2 - \left|b_4^+\right|^2\end{array}$ & $\left\{c;y;-\right\}$ & \\ [+9pt]
$I_0P_z^\odot$ & $ \begin{array}{l}\left|{\cal M}_1^-\right|^2 - \left|{\cal 
M}_1^+\right|^2 + \left|{\cal M}_2^-\right|^2 - \left|{\cal 
M}_2^+\right|^2 \\- \left|{\cal M}_3^-\right|^2 + \left|{\cal 
M}_3^+\right|^2 - \left|{\cal M}_4^-\right|^2 + \left|{\cal 
M}_4^+\right|^2\end{array}$ & 
$ 2\Re\left(b_1^-b_3^{-*} - b_1^+b_3^{+*} + 
b_2^-b_4^{-*} - b_2^+b_4^{+*}\right)$ & $\left\{c;z;-\right\}$ &  \\ [+9pt]
$I_0P^\odot_{x^\prime}$ & $ 2\Re\left({\cal M}_1^-{\cal M}_2^{-*} - 
{\cal M}_1^+{\cal M}_2^{+*} + 
{\cal M}_3^-{\cal M}_4^{-*} -{\cal M}_3^+{\cal M}_4^{+*}\right)$ & 
$ -2\Im\left(b_1^-b_2^{-*} - b_1^+b_2^{+*} +
b_3^-b_4^{-*} - b_3^+b_4^{+*}\right)$ & $\left\{c;-;x^\prime\right\}$ & B$_\odot$R \\ [+9pt]
$I_0P^\odot_{y^\prime}$ & $ -2\Im\left({\cal M}_1^-{\cal M}_2^{-*} - 
{\cal M}_1^+{\cal M}_2^{+*} + 
{\cal M}_3^-{\cal M}_4^{-*} -{\cal M}_3^+{\cal M}_4^{+*}\right)$ & 
$ \begin{array}{l}-\left|b_1^-\right|^2 + \left|b_1^+\right|^2
 + \left|b_2^-\right|^2 - \left|b_2^+\right|^2 \\- \left|b_3^-\right|^2 
 + \left|b_3^+\right|^2 + \left|b_4^-\right|^2 - \left|b_4^+\right|^2\end{array}$ & $\left\{c;-;y^\prime\right\}$ & \\ [+9pt]
$I_0P^\odot_{z^\prime}$ & $\begin{array}{l} -\left|{\cal M}_1^-\right|^2 + \left|{\cal 
M}_1^+\right|^2 + \left|{\cal M}_2^-\right|^2 - \left|{\cal 
M}_2^+\right|^2 \\- \left|{\cal M}_3^-\right|^2 + \left|{\cal 
M}_3^+\right|^2 + \left|{\cal M}_4^-\right|^2 - \left|{\cal 
M}_4^+\right|^2\end{array}$ & 
$ 2\Re\left(b_1^-b_2^{-*} - b_1^+b_2^{+*} + 
b_3^-b_4^{-*} - b_3^+b_4^{+*}\right)$ & $\left\{c;-;z^\prime\right\}$ &  \\ [+9pt]
$I_0{\cal O}^\odot_{xx^\prime}$ & $2\Re\left({\cal M}_2^-{\cal M}_3^{-*} - 
{\cal M}_2^+{\cal M}_3^{+*} + 
{\cal M}_1^-{\cal M}_4^{-*} -{\cal M}_1^+{\cal M}_4^{+*}\right)$ & 
$ 2\Re\left(b_2^-b_3^{-*} - 
     b_2^+b_3^{+*} - b_1^-b_4^{-*} + b_1^+b_4^{+*}\right)$ & $\left\{c;x;x^\prime\right\}$ & B$_\odot$TR \\ [+9pt]
$I_0{\cal O}^\odot_{xy^\prime}$ & $2\Im\left({\cal M}_2^-{\cal M}_3^{-*} - 
{\cal M}_2^+{\cal M}_3^{+*} - 
{\cal M}_1^-{\cal M}_4^{-*} +{\cal M}_1^+{\cal M}_4^{+*}\right)$ & 
$ 2\Im\left(b_1^-b_3^{-*} - 
      b_1^+b_3^{+*}  - b_2^-b_4^{-*} + b_2^+b_4^{+*}\right)$ &  $\left\{c;x;y^\prime\right\}$ & \\ [+9pt]
$I_0{\cal O}^\odot_{xz^\prime}$ & $2\Re\left(-{\cal M}_1^-{\cal M}_3^{-*} + 
{\cal M}_1^+{\cal M}_3^{+*} + 
{\cal M}_2^-{\cal M}_4^{-*} -{\cal M}_2^+{\cal M}_4^{+*}\right)$ & 
$ -2\Im\left( b_2^-b_3^{-*} - 
      b_2^+b_3^{+*} + b_1^-b_4^{-*} - b_1^+b_4^{+*}\right)$ & $\left\{c;x;z^\prime\right\}$ & \\ [+9pt]
$I_0{\cal O}^\odot_{yx^\prime}$ & $-2\Im\left({\cal M}_2^-{\cal M}_3^{-*} - 
{\cal M}_2^+{\cal M}_3^{+*} + 
{\cal M}_1^-{\cal M}_4^{-*} -{\cal M}_1^+{\cal M}_4^{+*}\right)$ & 
$ -2\Im\left(b_1^-b_2^{-*} - 
      b_1^+b_2^{+*} - b_3^-b_4^{-*} + b_3^+b_4^{+*}\right)$ & $\left\{c;y;x^\prime\right\}$ & \\ [+9pt]
$I_0{\cal O}^\odot_{yy^\prime}$ & $2\Re\left({\cal M}_2^-{\cal M}_3^{-*} - 
{\cal M}_2^+{\cal M}_3^{+*} - 
{\cal M}_1^-{\cal M}_4^{-*} +{\cal M}_1^+{\cal M}_4^{+*}\right)$ & 
$ \begin{array}{l}-\left|b_1^-\right|^2 + \left|b_1^+\right|^2 + \left|b_2^-\right|^2 -
   \left|b_2^+\right|^2 \\+ \left|b_3^-\right|^2 - \left|b_3^+\right|^2 - 
   \left|b_4^-\right|^2 + \left|b_4^+\right|^2\end{array}$ & $\left\{c;y;y^\prime\right\}$ & \\ [+9pt]
$I_0{\cal O}^\odot_{yz^\prime}$ & $2\Im\left({\cal M}_1^-{\cal M}_3^{-*} -
{\cal M}_1^+{\cal M}_3^{+*} - 
{\cal M}_2^-{\cal M}_4^{-*} +{\cal M}_2^+{\cal M}_4^{+*}\right)$ & 
$ 2\Re\left(b_1^-b_2^{-*} - 
     b_1^+b_2^{+*} - b_3^-b_4^{-*} + b_3^+b_4^{+*}\right)$ & $\left\{c;y;z^\prime\right\}$ & \\ [+9pt]
$I_0{\cal O}^\odot_{zx^\prime}$ & $2\Re\left(-{\cal M}_1^-{\cal M}_2^{-*} + 
{\cal M}_1^+{\cal M}_2^{+*} + 
{\cal M}_3^-{\cal M}_4^{-*} -{\cal M}_3^+{\cal M}_4^{+*}\right)$ & 
$ -2\Im\left(b_2^-b_3^{-*} - 
      b_2^+b_3^{+*} -b_1^-b_4^{-*} + b_1^+b_4^{+*}\right)$ & $\left\{c;z;x^\prime\right\}$ & \\ [+9pt]
$I_0{\cal O}^\odot_{zy^\prime}$ & $2\Im\left({\cal M}_1^-{\cal M}_2^{-*} - 
{\cal M}_1^+{\cal M}_2^{+*} - 
{\cal M}_3^-{\cal M}_4^{-*} +{\cal M}_3^+{\cal M}_4^{+*}\right)$ & 
$ 2\Re\left(b_1^-b_3^{-*} - 
     b_1^+b_3^{+*} - b_2^-b_4^{-*} + b_2^+b_4^{+*}\right)$ & $\left\{c;z;y^\prime\right\}$ & \\ [+9pt]
$I_0{\cal O}^\odot_{zz^\prime}$ & $ \begin{array}{l}\left|{\cal M}_1^-\right|^2 - \left|{\cal 
M}_1^+\right|^2 - \left|{\cal M}_2^-\right|^2 + \left|{\cal 
M}_2^+\right|^2 \\- \left|{\cal M}_3^-\right|^2 + \left|{\cal 
M}_3^+\right|^2 + \left|{\cal M}_4^-\right|^2 - \left|{\cal 
M}_4^+\right|^2\end{array}$ & $ 2\Re\left(- b_2^-b_3^{-*} + 
     b_2^+b_3^{+*} - b_1^-b_4^{-*} + b_1^+b_4^{+*}\right)$ & $\left\{c;z;z^\prime\right\}$ & \\ [+9pt]
\hline 
\end{tabular}
\end{center} 
\end{table}

\begin{table}[p]
\caption{Polarization observables of single and double pion photoproduction in 
terms of the helicity and transversity amplitudes. These observables
arise with linearly polarized photons, and are proportional to 
$\sin{2\beta}$ in the cross section. The notation is as in table 
\ref{tab:doublepolobs}.\label{tab:doublepolobsls}}
\begin{center}
\begin{tabular}{c|l|l|c|c} 
\hline Obs. & Helicity Form & Transversity Form & Expt. & Type \\ 
\hline 
$I_0I^s$ & $ -2\Im\left({\cal M}_1^+{\cal M}_1^{-*} + 
{\cal M}_2^+{\cal M}_2^{-*} + 
{\cal M}_3^+{\cal M}_3^{-*} +{\cal M}_4^+{\cal M}_4^{-*}\right)$ & 
$ -2\Im\left(b_1^+b_1^{-*} +b_2^+b_2^{-*} 
     + b_3^+b_3^{-*}  + b_4^+b_4^{-*} \right)$ & 
$\left\{L\left(\pm\frac{\pi}{4}\right);-;-\right\}$ & B$_\ell$ \\ 
$I_0P_x^s$ & $ -2\Im\left({\cal M}_1^+{\cal M}_3^{-*} - 
{\cal M}_1^-{\cal M}_3^{+*} + 
{\cal M}_2^+{\cal M}_4^{-*} -{\cal M}_2^-{\cal M}_4^{+*}\right)$ & 
$ 2\Re\left(-b_1^+b_3^{-*} + b_1^-b_3^{+*}- b_2^+b_4^{-*} 
+ b_2^-b_4^{+*}\right)$ & 
$\left\{L\left(\pm\frac{\pi}{4}\right);x;-\right\}$ &  B$_\ell$T  \\ 
$I_0P_y^s$ & $ 2\Re\left(-{\cal M}_1^+{\cal M}_3^{-*} + 
{\cal M}_1^-{\cal M}_3^{+*} - 
{\cal M}_2^+{\cal M}_4^{-*} +{\cal M}_2^-{\cal M}_4^{+*}\right)$ & 
$ -2\Im\left(b_1^+b_1^{-*} + b_2^+b_2^{-*} - b_3^+b_3^{-*} -
 b_4^+b_4^{-*}\right)$ & $\left\{L\left(\pm\frac{\pi}{4}\right);y;-\right\}$ & \\ 
$I_0P_z^s$ & $ 2\Im\left({\cal M}_1^+{\cal M}_1^{-*}+ 
{\cal M}_2^+{\cal M}_2^{-*} - 
{\cal M}_3^+{\cal M}_3^{-*} +{\cal M}_4^+{\cal M}_4^{-*}\right)$ & 
$ 2\Im\left(b_1^+b_3^{-*} - b_1^-b_3^{+*} + 
b_2^+b_4^{-*} - b_2^-b_4^{+*}\right)$ & 
$\left\{L\left(\pm\frac{\pi}{4}\right);z;-\right\}$ &  \\ 
$I_0P^s_{x^\prime}$ & $ 2\Im\left({\cal M}_1^+{\cal M}_2^{-*} - 
{\cal M}_1^-{\cal M}_2^{+*} + 
{\cal M}_3^+{\cal M}_4^{-*} -{\cal M}_3^-{\cal M}_4^{+*}\right)$ & 
$ 2\Re\left(b_1^+b_2^{-*} - b_1^-b_2^{+*} +
 b_3^+b_4^{-*} - b_3^-b_4^{+*}\right)$ & 
$\left\{L\left(\pm\frac{\pi}{4}\right);-;x^\prime\right\}$ & B$_\ell$R \\ 
$I_0P^s_{y^\prime}$ & $ 2\Re\left({\cal M}_1^+{\cal M}_2^{-*} - 
{\cal M}_1^-{\cal M}_2^{+*} + 
{\cal M}_3^+{\cal M}_4^{-*} -{\cal M}_3^-{\cal M}_4^{+*}\right)$ & 
$ -2\Im\left(b_1^+b_1^{-*} - 
      b_2^+b_2^{-*} + b_3^+b_3^{-*} -  b_4^+b_4^{-*} \right)$ & $\left\{L\left(\pm\frac{\pi}{4}\right);-;y^\prime\right\}$ &  \\ 
$I_0P^s_{z^\prime}$ & $ -2\Im\left({\cal M}_1^+{\cal M}_1^{-*} - 
{\cal M}_2^+{\cal M}_2^{-*} + 
{\cal M}_3^+{\cal M}_3^{-*} -{\cal M}_4^+{\cal M}_4^{-*}\right)$ & 
$ 2\Im\left(b_1^+b_2^{-*} - b_1^-b_2^{+*} + 
b_3^+b_4^{-*} - b_3^-b_4^{+*}\right)$ & 
$\left\{L\left(\pm\frac{\pi}{4}\right);-;z^\prime\right\}$ &  \\ 
$I_0{\cal O}^s_{xx^\prime}$ & $ 2\Im\left({\cal M}_2^+{\cal M}_3^{-*} - 
{\cal M}_2^-{\cal M}_3^{+*} + 
{\cal M}_1^+{\cal M}_4^{-*} -{\cal M}_1^-{\cal M}_4^{+*}\right)$ & 
$ 2\Im\left(b_2^+b_3^{-*} - b_2^-b_3^{+*} -
 b_1^+b_4^{-*} + b_1^-b_4^{+*}\right)$ & 
$\left\{L\left(\pm\frac{\pi}{4}\right);x;x^\prime\right\}$ &  B$_\ell$TR\\ 
$I_0{\cal O}^s_{xy^\prime}$ & $ 2\Re\left(-{\cal M}_2^+{\cal M}_3^{-*} + 
{\cal M}_2^-{\cal M}_3^{+*} + 
{\cal M}_1^+{\cal M}_4^{-*} -{\cal M}_1^-{\cal M}_4^{+*}\right)$ & 
$ 2\Re\left(- b_1^+b_3^{-*} + b_1^-b_3^{+*} + 
     b_2^+b_4^{-*} - b_2^-b_4^{+*}\right)$ & $\left\{L\left(\pm\frac{\pi}{4}\right);x;y^\prime\right\}$ &  \\ 
$I_0{\cal O}^s_{xz^\prime}$ & $ -2\Im\left({\cal M}_1^+{\cal M}_3^{-*} - 
{\cal M}_1^-{\cal M}_3^{+*} - 
{\cal M}_2^+{\cal M}_4^{-*} +{\cal M}_2^-{\cal M}_4^{+*}\right)$ & 
$ 2\Re\left(b_2^+b_3^{-*} - b_2^-b_3^{+*} +
 b_1^+b_4^{-*} - b_1^-b_4^{+*}\right)$ & $\left\{L\left(\pm\frac{\pi}{4}\right);x;z^\prime\right\}$ &  \\ 
$I_0{\cal O}^s_{yx^\prime}$ & $ 2\Re\left({\cal M}_2^+{\cal M}_3^{-*} - 
{\cal M}_2^-{\cal M}_3^{+*} + 
{\cal M}_1^+{\cal M}_4^{-*} -{\cal M}_1^-{\cal M}_4^{+*}\right)$ & 
$ 2\Re\left(b_1^+b_2^{-*} - b_1^-b_2^{+*} -
 b_3^+b_4^{-*} + b_3^-b_4^{+*}\right)$ & $\left\{L\left(\pm\frac{\pi}{4}\right);y;x^\prime\right\}$ &  \\ 
$I_0{\cal O}^s_{yy^\prime}$ & $ 2\Im\left({\cal M}_2^+{\cal M}_3^{-*} - 
{\cal M}_2^-{\cal M}_3^{+*} - 
{\cal M}_1^+{\cal M}_4^{-*} +{\cal M}_1^-{\cal M}_4^{+*}\right)$ & 
$ -2\Im\left(b_1^+b_1^{-*}  -b_2^+b_2^{-*} 
 - b_3^+b_3^{-*} +  b_4^+b_4^{-*} \right)$ & $\left\{L\left(\pm\frac{\pi}{4}\right);y;y^\prime\right\}$ &  \\ 
$I_0{\cal O}^s_{yz^\prime}$ & $ 2\Re\left(-{\cal M}_1^+{\cal M}_3^{-*} + 
{\cal M}_1^-{\cal M}_3^{+*} + 
{\cal M}_2^+{\cal M}_4^{-*} -{\cal M}_2^-{\cal M}_4^{+*}\right)$ & 
$ 2\Im\left(b_1^+b_2^{-*} - b_1^-b_2^{+*} 
     - b_3^+b_4^{-*} + b_3^-b_4^{+*}\right)$ & $\left\{L\left(\pm\frac{\pi}{4}\right);y;z^\prime\right\}$ &  \\ 
$I_0{\cal O}^s_{zx^\prime}$ & $ -2\Im\left({\cal M}_1^+{\cal M}_2^{-*} - 
{\cal M}_1^-{\cal M}_2^{+*} - 
{\cal M}_3^+{\cal M}_4^{-*} +{\cal M}_3^-{\cal M}_4^{+*}\right)$ & 
$ 2\Re\left(b_2^+b_3^{-*} - b_2^-b_3^{+*} -
 b_1^+b_4^{-*} + b_1^-b_4^{+*}\right)$ & $\left\{L\left(\pm\frac{\pi}{4}\right);z;x^\prime\right\}$ &  \\ 
$I_0{\cal O}^s_{zy^\prime}$ & $ 2\Re\left(-{\cal M}_1^+{\cal M}_2^{-*} + 
{\cal M}_1^-{\cal M}_2^{+*} + 
{\cal M}_3^+{\cal M}_4^{-*} -{\cal M}_3^-{\cal M}_4^{+*}\right)$ & 
$ 2\Im\left( b_1^+b_3^{-*} - b_1^-b_3^{+*}
    - b_2^+b_4^{-*} +  b_2^-b_4^{+*}\right)$ & $\left\{L\left(\pm\frac{\pi}{4}\right);z;y^\prime\right\}$ &  \\ 
$I_0{\cal O}^s_{zz^\prime}$ & $ 2\Im\left({\cal M}_1^+{\cal M}_1^{-*} - 
{\cal M}_2^+{\cal M}_2^{-*} - 
{\cal M}_3^+{\cal M}_3^{-*} +{\cal M}_4^+{\cal M}_4^{-*} \right)$ & 
$ -2\Im\left(b_2^+b_3^{-*} - b_2^-b_3^{+*} + 
b_1^+b_4^{-*} - b_1^-b_4^{+*}\right)$ & $\left\{L\left(\pm\frac{\pi}{4}\right);z;z^\prime\right\}$ &  \\ 
\hline 
\end{tabular}
\end{center} 
\end{table}

\begin{table}[p]
\caption{Polarization observables of single and double pion photoproduction in 
terms of the helicity and transversity amplitudes. These observables
arise with linearly polarized photons, and are proportional to 
$\cos{2\beta}$ in the cross section. The notation is as in table 
\ref{tab:doublepolobs}.\label{tab:doublepolobslc}}
\begin{center}
\begin{tabular}{c|l|l|c|c} 
\hline Obs. & Helicity Form & Transversity Form & Expt. & Type \\ 
\hline 
$I_0I^c$ & $ -2\Re\left({\cal M}_1^+{\cal M}_1^{-*} + 
{\cal M}_2^+{\cal M}_2^{-*} + 
{\cal M}_3^+{\cal M}_3^{-*} + {\cal M}_4^+{\cal M}_4^{-*}\right)$ & 
$ 2\Re\left(-(b_1^+b_1^{-*})  - b_2^+b_2^{-*} 
- b_3^+b_3^{-*} -  b_4^+b_4^{-*} \right)$ & 
$\left\{L\left(\frac{\pi}{2},0\right);-;-\right\}$ & B$_\ell$ \\ 
$I_0P_x^c$ & $ -2\Re\left({\cal M}_1^+{\cal M}_3^{-*} + 
{\cal M}_1^-{\cal M}_3^{+*} + 
{\cal M}_2^+{\cal M}_4^{-*} +{\cal M}_2^-{\cal M}_4^{+*}\right)$ & 
$ 2\Im\left( b_1^+b_3^{-*} + b_1^-b_3^{+*} +
 b_2^+b_4^{-*} + b_2^-b_4^{+*}\right)$ & 
$\left\{L\left(\frac{\pi}{2},0\right);x;-\right\}$ & B$_\ell$T \\ 
$I_0P_y^c$ & $ 2\Im\left({\cal M}_1^+{\cal M}_3^{-*} + 
{\cal M}_1^-{\cal M}_3^{+*} + 
{\cal M}_2^+{\cal M}_4^{-*} +{\cal M}_2^-{\cal M}_4^{+*}\right)$ & 
$ 2\Re\left(-(b_1^+b_1^{-*}) - b_2^+b_2^{-*} 
 + b_3^+b_3^{-*} + b_4^+b_4^{-*} \right)$ & $\left\{L\left(\frac{\pi}{2},0\right);y;-\right\}$ &  \\
$I_0P_z^c$ & $ 2\Re\left({\cal M}_1^+{\cal M}_1^{-*} + 
{\cal M}_2^+{\cal M}_2^{-*} - 
{\cal M}_3^+{\cal M}_3^{-*} -{\cal M}_4^+{\cal M}_4^{-*} \right)$ & 
$ 2\Re\left( b_1^+b_3^{-*} + 
     b_1^-b_3^{+*} +  b_2^+b_4^{-*} + b_2^-b_4^{+*}\right)$ & 
$\left\{L\left(\frac{\pi}{2},0\right);z;-\right\}$ & \\ 
$I_0P^c_{x^\prime}$ & $ 2\Re\left({\cal M}_1^+{\cal M}_2^{-*} + 
{\cal M}_1^-{\cal M}_2^{+*} + 
{\cal M}_3^+{\cal M}_4^{-*} +{\cal M}_3^-{\cal M}_4^{+*}\right)$ & 
$ -2\Im\left(b_1^+b_2^{-*} + b_1^-b_2^{+*} + 
b_3^+b_4^{-*} + b_3^-b_4^{+*}\right)$ & 
$\left\{L\left(\frac{\pi}{2},0\right);-;x^\prime\right\}$ & B$_\ell$R  \\ 
$I_0P^c_{y^\prime}$ & $ -2\Im\left({\cal M}_1^+{\cal M}_2^{-*} + 
{\cal M}_1^-{\cal M}_2^{+*} + 
{\cal M}_3^+{\cal M}_4^{-*} +{\cal M}_3^-{\cal M}_4^{+*}\right)$ & 
$ 2\Re\left(-(b_1^+b_1^{-*})  + 
     b_2^+b_2^{-*} - b_3^+b_3^{-*} + b_4^+b_4^{-*} \right)$ & $\left\{L\left(\frac{\pi}{2},0\right);-;y^\prime\right\}$ & \\ 
$I_0P^c_{z^\prime}$ & $ 2\Re\left(-{\cal M}_1^+{\cal M}_1^{-*} + 
{\cal M}_2^+{\cal M}_2^{-*} - 
{\cal M}_3^+{\cal M}_3^{-*} +{\cal M}_4^+{\cal M}_4^{-*}\right)$ & 
$ 2\Re\left(b_1^+b_2^{-*} + b_1^-b_2^{+*} +
 b_3^+b_4^{-*} + b_3^-b_4^{+*}\right)$ & 
$\left\{L\left(\frac{\pi}{2},0\right);-;z^\prime\right\}$ & \\ 
$I_0{\cal O}^c_{xx^\prime}$ & $ 2\Re\left({\cal M}_2^+{\cal M}_3^{-*} + 
{\cal M}_2^-{\cal M}_3^{+*} + 
{\cal M}_1^+{\cal M}_4^{-*} +{\cal M}_1^-{\cal M}_4^{+*}\right)$ & 
$ 2\Re\left(b_2^+b_3^{-*} + b_2^-b_3^{+*} -
 b_1^+b_4^{-*} - b_1^-b_4^{+*}\right)$ & 
$\left\{L\left(\frac{\pi}{2},0\right);x;x^\prime\right\}$ & B$_\ell$TR  \\ 
$I_0{\cal O}^c_{xy^\prime}$ & $ 2\Im\left({\cal M}_2^+{\cal M}_3^{-*} + 
{\cal M}_2^-{\cal M}_3^{+*} - 
{\cal M}_1^+{\cal M}_4^{-*} -{\cal M}_1^-{\cal M}_4^{+*}\right)$ & 
$ 2\Im\left(b_1^+b_3^{-*} + b_1^-b_3^{+*} -
 b_2^+b_4^{-*} - b_2^-b_4^{+*}\right)$ & $\left\{L\left(\frac{\pi}{2},0\right);x;y^\prime\right\}$ &  \\ 
$I_0{\cal O}^c_{xz^\prime}$ & $ 2\Re\left(-{\cal M}_1^+{\cal M}_3^{-*} - 
{\cal M}_1^-{\cal M}_3^{+*} + 
{\cal M}_2^+{\cal M}_4^{-*} +{\cal M}_2^-{\cal M}_4^{+*}\right)$ & 
$ -2\Im\left(b_2^+b_3^{-*} + b_2^-b_3^{+*} +
 b_1^+b_4^{-*} + b_1^-b_4^{+*}\right)$ & $\left\{L\left(\frac{\pi}{2},0\right);x;z^\prime\right\}$ &  \\ 
$I_0{\cal O}^c_{yx^\prime}$ & $ -2\Im\left({\cal M}_2^+{\cal M}_3^{-*} + 
{\cal M}_2^-{\cal M}_3^{+*} + 
{\cal M}_1^+{\cal M}_4^{-*} +{\cal M}_1^-{\cal M}_4^{+*}\right)$ & 
$ -2\Im\left( b_1^+b_2^{-*} + b_1^-b_2^{+*}  -
 b_3^+b_4^{-*} -  b_3^-b_4^{+*}\right)$ & $\left\{L\left(\frac{\pi}{2},0\right);x;x^\prime\right\}$ &  \\ 
$I_0{\cal O}^c_{yy^\prime}$ & $ 2\Re\left({\cal M}_2^+{\cal M}_3^{-*} + 
{\cal M}_2^-{\cal M}_3^{+*} - 
{\cal M}_1^+{\cal M}_4^{-*} -{\cal M}_1^-{\cal M}_4^{+*}\right)$ & 
$ 2\Re\left(-(b_1^+b_1^{-*}) + b_2^+b_2^{-*} + 
     b_3^+b_3^{-*}- b_4^+b_4^{-*} \right)$ & $\left\{L\left(\frac{\pi}{2},0\right);x;y^\prime\right\}$ &  \\ 
$I_0{\cal O}^c_{yz^\prime}$ & $ 2\Im\left({\cal M}_1^+{\cal M}_3^{-*} + 
{\cal M}_1^-{\cal M}_3^{+*} -
{\cal M}_2^+{\cal M}_4^{-*} -{\cal M}_2^-{\cal M}_4^{+*}\right)$ & 
$ 2\Re\left(b_1^+b_2^{-*} + b_1^-b_2^{+*}  -
 b_3^+b_4^{-*} - b_3^-b_4^{+*}\right)$ & $\left\{L\left(\frac{\pi}{2},0\right);x;z^\prime\right\}$ &  \\ 
$I_0{\cal O}^c_{zx^\prime}$ & $ 2\Re\left(-{\cal M}_1^+{\cal M}_2^{-*} - 
{\cal M}_1^-{\cal M}_2^{+*} + 
{\cal M}_3^+{\cal M}_4^{-*} +{\cal M}_3^-{\cal M}_4^{+*}\right)$ & 
$ 2\Im\left(-b_2^+b_3^{-*} - b_2^-b_3^{+*} +
 b_1^+b_4^{-*} + b_1^-b_4^{+*}\right)$ & $\left\{L\left(\frac{\pi}{2},0\right);z;x^\prime\right\}$ &  \\ 
$I_0{\cal O}^c_{zy^\prime}$ & $ 2\Im\left({\cal M}_1^+{\cal M}_2^{-*} + 
{\cal M}_1^-{\cal M}_2^{+*} - 
{\cal M}_3^+{\cal M}_4^{-*} -{\cal M}_3^-{\cal M}_4^{+*}\right)$ & 
$ 2\Re\left(b_1^+b_3^{-*} + b_1^-b_3^{+*} -
 b_2^+b_4^{-*} - b_2^-b_4^{+*}\right)$ & $\left\{L\left(\frac{\pi}{2},0\right);z;y^\prime\right\}$ &  \\ 
$I_0{\cal O}^c_{zz^\prime}$ & $ 2\Re\left({\cal M}_1^+{\cal M}_1^{-*} - 
{\cal M}_2^+{\cal M}_2^{-*} - 
{\cal M}_3^+{\cal M}_3^{-*} + {\cal M}_4^+{\cal M}_4^{-*}\right)$ & 
$ 2\Re\left(- b_2^+b_3^{-*} - b_2^-b_3^{+*} -
 b_1^+b_4^{-*} - b_1^-b_4^{+*}\right)$ & 
$\left\{L\left(\frac{\pi}{2},0\right);z;z^\prime\right\}$ &  \\ 
\hline 
\end{tabular}
\end{center} 
\end{table}

As was the case with the pion-induced reactions, there are a number of relationships among these
64 polarization observables. In fact, there are 28 relations that arise from consideration of the
absolute magnitudes of the helicity or transversity amplitudes, and another 21 that arise from
consideration of their phases, leaving 15 independent quantities. We list here the relations that
arise from considerations of the absolute magnitudes of the amplitudes $b_i$.

\beqy
&&\left[P_{x^\prime}+\xi {\cal O}_{yx^\prime}+\zeta
\left(P_{x^\prime}^\odot+\xi{\cal O}_{yx^\prime}^\odot\right)\right]^2
+\left[P_{z^\prime}+\xi{\cal O}_{yz^\prime} +\zeta
\left(P_{z^\prime}^\odot+\xi{\cal O}_{yz^\prime}^\odot\right)\right]^2\nonumber\\[-3pt]
&&=\left[1+\xi P_y+\zeta\left(I^\odot+\xi P_y^\odot\right)\right]^2-
\left[P_{y^\prime}+\xi {\cal O}_{yy^\prime}+\zeta\left(P_{y^\prime}^\odot+\xi{\cal
O}_{yy^\prime}^\odot\right)\right]^2,\nonumber\\[+5pt]
&&\left[P_x+\xi{\cal O}_{xy^\prime} +\zeta
\left(P_x^\odot+\xi{\cal O}_{xy^\prime}^\odot\right)\right]^2
+\left[P_z+\xi{\cal O}_{zy^\prime} +\zeta
\left(P_z^\odot+\xi{\cal O}_{zy^\prime}^\odot\right)\right]^2\nonumber\\[-3pt]
&&=\left[1+\xi P_{y^\prime}+\zeta\left(I^\odot+\xi P_{y^\prime}^\odot\right)\right]^2-
\left[P_y+\xi {\cal O}_{yy^\prime}+\zeta\left(P_y^\odot+\xi{\cal
O}_{yy^\prime}^\odot\right)\right]^2,\nonumber\\[+5pt]
&&\left[{\cal O}_{xz^\prime}-\xi{\cal O}_{zx^\prime}+\zeta
\left({\cal O}_{xz^\prime}^\odot-\xi{\cal O}_{zx^\prime}^\odot\right)\right]^2
+\left[{\cal O}_{xx^\prime}+\xi{\cal O}_{zz^\prime}-\zeta
\left({\cal O}_{xx^\prime}^\odot+\xi{\cal O}_{zz^\prime}^\odot\right)\right]^2\nonumber\\[-3pt]
&&=\left[1+\xi{\cal O}_{yy^\prime} -\zeta\left(I^\odot+\xi{\cal
O}_{yy^\prime}^\odot\right)\right]^2-
\left[P_y+\xi P_{y^\prime}-\zeta\left(P_y^\odot+\xi P_{y^\prime}^\odot\right)\right]^2,\nonumber\\[+5pt]
&&\left[I^s+\xi{\cal O}_{yy^\prime}^s+\zeta\left(P_y^s+\xi P_{y^\prime}^s\right)\right]^2+
\left[I^c+\xi{\cal O}_{yy^\prime}^c+\zeta\left(P_y^c+\xi P_{y^\prime}^c\right)\right]^2\nonumber\\[-3pt]
&&=\left[1+\xi{\cal O}_{yy^\prime}+\zeta\left(P_y+\xi P_{y^\prime}\right)\right]^2-
\left[I^\odot+\xi{\cal O}_{yy^\prime}^\odot+\zeta\left(P_y^\odot+\xi P_{y^\prime}^\odot\right)\right]^2,\nonumber \\[+5pt]
&&\left[P_{x^\prime}^s+\xi{\cal O}_{yx^\prime}^s+\zeta\left(P_{z^\prime}^c+\xi{\cal O}_{yz^\prime}^c\right)\right]^2+
\left[P_{z^\prime}^s+\xi{\cal O}_{yz^\prime}^s-\zeta\left(P_{x^\prime}^c+\xi{\cal O}_{yx^\prime}^c\right)\right]^2\nonumber\\[-3pt]
&&=\left[1+\xi P_y+\zeta\left(P_{y^\prime}^\odot+\xi{\cal O}_{yy^\prime}^\odot\right)\right]^2-
\left[I^\odot+\xi P_y^\odot+\zeta\left(P_{y^\prime}+\xi{\cal O}_{yy^\prime}\right)\right]^2,\nonumber \\[+5pt]
&&\left[P_x^s+\xi{\cal O}_{xy^\prime}^s+\zeta\left(P_z^c+\xi{\cal O}_{zy^\prime}^c\right)\right]^2+
\left[P_z^s+\xi{\cal O}_{zy^\prime}^s-\zeta\left(P_x^c+\xi{\cal O}_{xy^\prime}^c\right)\right]^2\nonumber\\[-3pt]
&&=\left[1+\xi P_{y^\prime}-\zeta\left(P_y^\odot+\xi{\cal O}_{yy^\prime}^\odot\right)\right]^2-
\left[I^\odot+\xi P_{y^\prime}^\odot-\zeta\left(P_y+\xi{\cal O}_{yy^\prime}\right)\right]^2,\nonumber\\[+5pt]
&&\left[{\cal O}_{xx^\prime}^s+\xi {\cal O}_{zz^\prime}^s+\zeta
\left({\cal O}_{xz^\prime}^c-\xi{\cal O}_{zx^\prime}^c\right)\right]^2
+\left[{\cal O}_{xz^\prime}^s-\xi {\cal O}_{zx^\prime}^s-\zeta
\left({\cal O}_{xx^\prime}^c+\xi{\cal O}_{zz^\prime}^c\right)\right]^2\nonumber\\[-3pt]
&&=\left[1+\xi{\cal O}_{yy^\prime}+\zeta\left(P_{y^\prime}^\odot+\xi P_y^\odot\right)\right]^2-
\left[I^\odot+\xi{\cal O}_{yy^\prime}^\odot+\zeta\left(P_{y^\prime}+\xi
P_y\right)\right]^2.
\eeqy
In each of these equations, $\xi$ and $\zeta$ can independently take either of
the values $\pm 1$, meaning that the seven equations shown above actually 
represent 28 identities. We can also obtain another 21 identities from considering
the phases of the transversity amplitudes, but we do not display these here. We
also point out that the equations above were obtained from considering the
transversity amplitudes. Had we considered the helicity amplitudes instead, we
would obtain a different set of 28 identities among the observables from the
magnitudes of the amplitudes, and another 21 identities from their phases. In
either case, we are left with 15 independent observables.

As was done in the case of $\pi N\to\pi\pi N$, we can use the identities above to write a number of
inequalities that the polarization observables for $\gamma N\to\pi\pi N$ must satisfy. These inequalities
are
\beqy
\left|1+\xi P_y+\zeta\left(I^\odot+\xi P_y^\odot\right)\right|&\ge&
\left\{\left|P_{y^\prime}+\xi {\cal O}_{yy^\prime}+\zeta\left(P_{y^\prime}^\odot+\xi{\cal
O}_{yy^\prime}^\odot\right)\right|,\right.\nonumber\\[-3pt]
\left|P_{x^\prime}+\xi {\cal O}_{yx^\prime}+\zeta
\left(P_{x^\prime}^\odot+\xi{\cal O}_{yx^\prime}^\odot\right)\right|,&&\left.
\left|P_{z^\prime}+\xi{\cal O}_{yz^\prime} +\zeta
\left(P_{z^\prime}^\odot+\xi{\cal O}_{yz^\prime}^\odot\right)\right|\right\},\nonumber\\[+5pt]
\left|1+\xi P_{y^\prime}+\zeta\left(I^\odot+\xi P_{y^\prime}^\odot\right)\right|&\ge&
\left\{\left|P_y+\xi {\cal O}_{yy^\prime}+\zeta\left(P_y^\odot+\xi{\cal
O}_{yy^\prime}^\odot\right)\right|\right.,\nonumber\\[-3pt]
\left|P_x+\xi{\cal O}_{xy^\prime} +\zeta
\left(P_x^\odot+\xi{\cal O}_{xy^\prime}^\odot\right)\right|,&&\left.
\left|P_z+\xi{\cal O}_{zy^\prime} +\zeta
\left(P_z^\odot+\xi{\cal O}_{zy^\prime}^\odot\right)\right|\right\},\nonumber\\[+5pt]
\left|1+\xi{\cal O}_{yy^\prime} -\zeta\left(I^\odot+\xi{\cal
O}_{yy^\prime}^\odot\right)\right|&\ge&
\left\{\left|P_y+\xi P_{y^\prime}-\zeta\left(P_y^\odot+\xi
P_{y^\prime}^\odot\right)\right|\right.,\nonumber\\[-3pt]
\left|{\cal O}_{xz^\prime}-\xi{\cal O}_{zx^\prime}+\zeta
\left({\cal O}_{xz^\prime}^\odot-\xi{\cal O}_{zx^\prime}^\odot\right)\right|,&&\left.
\left|{\cal O}_{xx^\prime}+\xi{\cal O}_{zz^\prime}-\zeta
\left({\cal O}_{xx^\prime}^\odot+\xi{\cal
O}_{zz^\prime}^\odot\right)\right|\right\},\nonumber\\[+5pt]
\left|1+\xi{\cal O}_{yy^\prime}+\zeta\left(P_y+\xi P_{y^\prime}\right)\right|&\ge&
\left\{\left|I^\odot+\xi{\cal O}_{yy^\prime}^\odot+\zeta\left(P_y^\odot+\xi
P_{y^\prime}^\odot\right)\right|\right.,\nonumber \\[-3pt]
\left|I^s+\xi{\cal O}_{yy^\prime}^s+\zeta\left(P_y^s+\xi
P_{y^\prime}^s\right)\right|,&&\left.
\left|I^c+\xi{\cal O}_{yy^\prime}^c+\zeta\left(P_y^c+
\xi P_{y^\prime}^c\right)\right|\right\},\nonumber\\[+5pt]
\left|1+\xi P_y+\zeta\left(P_{y^\prime}^\odot+\xi{\cal O}_{yy^\prime}^\odot\right)\right|&\ge&
\left\{\left|I^\odot+\xi P_y^\odot+\zeta\left(P_{y^\prime}+\xi{\cal O}_{yy^\prime}\right)\right|\right.,\nonumber \\[-3pt]
\left|P_{x^\prime}^s+\xi{\cal O}_{yx^\prime}^s+\zeta\left(P_{z^\prime}^c+\xi{\cal O}_{yz^\prime}^c\right)\right|,&&
\left.\left|P_{z^\prime}^s+\xi{\cal O}_{yz^\prime}^s-\zeta\left(P_{x^\prime}^c+
\xi{\cal O}_{yx^\prime}^c\right)\right|\right\}\nonumber\\[+5pt]
\left|1+\xi P_{y^\prime}-\zeta\left(P_y^\odot+\xi{\cal O}_{yy^\prime}^\odot\right)\right|&\ge&
\left\{\left|I^\odot+\xi P_{y^\prime}^\odot-\zeta\left(P_y+\xi{\cal O}_{yy^\prime}\right)\right|\right.,\nonumber\\[-3pt]
\left|P_x^s+\xi{\cal O}_{xy^\prime}^s+\zeta\left(P_z^c+\xi{\cal O}_{zy^\prime}^c\right)\right|,&&\left.
\left|P_z^s+\xi{\cal O}_{zy^\prime}^s-\zeta\left(P_x^c+\xi{\cal O}_{xy^\prime}^c\right)\right|\right\},\nonumber\\[+5pt]
\left|1+\xi{\cal O}_{yy^\prime}+\zeta\left(P_{y^\prime}^\odot+\xi P_y^\odot\right)\right|&\ge&
\left\{\left|I^\odot+\xi{\cal O}_{yy^\prime}^\odot+\zeta\left(P_{y^\prime}+\xi
P_y\right)\right|\right.,\nonumber\\[-3pt]
\left|{\cal O}_{xx^\prime}^s+\xi {\cal O}_{zz^\prime}^s+\zeta
\left({\cal O}_{xz^\prime}^c-\xi{\cal O}_{zx^\prime}^c \vphantom{{\cal O}_{yy^\prime}^\odot}\right)\right|
,&&\left.\left|{\cal O}_{xz^\prime}^s-\xi {\cal O}_{zx^\prime}^s-\zeta
\left({\cal O}_{xx^\prime}^c+\xi{\cal O}_{zz^\prime}^c\vphantom{{\cal O}_{yy^\prime}^\odot}\right)\right|\right\}.
\eeqy
These inequalities can also be manipulated (as was done for $\pi N\to\pi\pi N$) to lead to
\beqy
1+ P_y^2+\left(I^\odot\right)^2+ \left(P_y^\odot\right)^2&\ge&
\left\{P_{y^\prime}^2+ {\cal O}_{yy^\prime}^2+\left(P_{y^\prime}^\odot\right)^2+\left({\cal
O}_{yy^\prime}^\odot\right)^2,\right.\nonumber\\[+3pt]
P_{x^\prime}^2+ {\cal O}_{yx^\prime}^2+
\left(P_{x^\prime}^\odot\right)^2+\left({\cal O}_{yx^\prime}^\odot\right)^2,&&\left.
P_{z^\prime}^2+{\cal O}_{yz^\prime}^2 +
\left(P_{z^\prime}^\odot\right)^2+\left({\cal O}_{yz^\prime}^\odot\right)^2\right\},\nonumber\\[+5pt]
1+ P_{y^\prime}^2+\left(I^\odot\right)^2+ \left(P_{y^\prime}^\odot\right)^2&\ge&
\left\{P_y^2+ {\cal O}_{yy^\prime}^2+\left(P_y^\odot\right)^2+\left({\cal
O}_{yy^\prime}^\odot\right)^2\right.,\nonumber\\[+3pt]
P_x^2+{\cal O}_{xy^\prime}^2 +
\left(P_x^\odot\right)^2+\left({\cal O}_{xy^\prime}^\odot\right)^2,&&\left.
P_z^2+{\cal O}_{zy^\prime}^2 +
\left(P_z^\odot\right)^2+\left({\cal O}_{zy^\prime}^\odot\right)^2\right\},\nonumber\\[+5pt]
1+{\cal O}_{yy^\prime}^2 +\left(I^\odot\right)^2+\left({\cal
O}_{yy^\prime}^\odot\right)^2&\ge&
\left\{P_y^2+ P_{y^\prime}^2+\left(P_y^\odot\right)^2+
\left(P_{y^\prime}^\odot\right)^2\right.,\nonumber\\[+3pt]
{\cal O}_{xz^\prime}^2+{\cal O}_{zx^\prime}^2+
\left({\cal O}_{xz^\prime}^\odot\right)^2+\left({\cal O}_{zx^\prime}^\odot\right)^2,&&\left.
{\cal O}_{xx^\prime}^2+{\cal O}_{zz^\prime}^2+
\left({\cal O}_{xx^\prime}^\odot\right)^2+\left({\cal
O}_{zz^\prime}^\odot\right)^2\right\},\nonumber\\[+5pt]
1+{\cal O}_{yy^\prime}^2+P_y^2+ P_{y^\prime}^2&\ge&
\left\{\left(I^\odot\right)^2+\left({\cal O}_{yy^\prime}^\odot\right)^2+\left(P_y^\odot\right)^2+
\left(P_{y^\prime}^\odot\right)^2\right.,\nonumber \\[+3pt]
\left(I^s\right)^2+\left({\cal O}_{yy^\prime}^s\right)^2+\left(P_y^s\right)^2+
\left(P_{y^\prime}^s\right)^2,&&\left.
\left(I^c\right)^2+\left({\cal O}_{yy^\prime}^c\right)^2+\left(P_y^c\right)^2+
 \left(P_{y^\prime}^c\right)^2\right\},\nonumber\\[+5pt]
1+ P_y^2+\left(P_{y^\prime}^\odot\right)^2+\left({\cal O}_{yy^\prime}^\odot\right)^2&\ge&
\left\{\left(I^\odot\right)^2+\left( P_y^\odot\right)^2+P_{y^\prime}^2+{\cal O}_{yy^\prime}^2\right.,\nonumber \\[+3pt]
\left(P_{x^\prime}^s\right)^2+\left({\cal O}_{yx^\prime}^s\right)^2+\left(P_{z^\prime}^c\right)^2+\left({\cal O}_{yz^\prime}^c\right)^2,&&
\left.\left(P_{z^\prime}^s\right)^2+\left({\cal O}_{yz^\prime}^s\right)^2+\left(P_{x^\prime}^c\right)^2+
\left({\cal O}_{yx^\prime}^c\right)^2\right\}\nonumber\\[+5pt]
1+ P_{y^\prime}^2+\left(P_y^\odot\right)^2+\left({\cal O}_{yy^\prime}^\odot\right)^2&\ge&
\left\{\left(I^\odot\right)^2+ \left(P_{y^\prime}^\odot\right)^2+P_y^2+{\cal O}_{yy^\prime}^2\right.,\nonumber\\[+3pt]
\left(P_x^s\right)^2+\left({\cal O}_{xy^\prime}^s\right)^2+\left(P_z^c\right)^2+\left({\cal O}_{zy^\prime}^c\right)^2,&&\left.
\left(P_z^s\right)^2+\left({\cal O}_{zy^\prime}^s\right)^2+\left(P_x^c\right)^2+\left({\cal O}_{xy^\prime}^c\right)^2\right\},\nonumber\\[+5pt]
1+{\cal O}_{yy^\prime}^2+\left(P_{y^\prime}^\odot\right)^2+ \left(P_y^\odot\right)^2&\ge&
\left\{\left(I^\odot\right)^2+\left({\cal O}_{yy^\prime}^\odot\right)^2+P_{y^\prime}^2+
P_y^2\right.,\nonumber\\[+3pt]
\left({\cal O}_{xx^\prime}^s\right)^2+ \left({\cal O}_{zz^\prime}^s\right)^2+
\left({\cal O}_{xz^\prime}^c\right)^2+\left({\cal O}_{zx^\prime}^c\right)^2 \vphantom{{\cal O}_{yy^\prime}^\odot}
,&&\left.\left({\cal O}_{xz^\prime}^s\right)^2+ \left({\cal O}_{zx^\prime}^s\right)^2+
\left({\cal O}_{xx^\prime}^c\right)^2+\left({\cal O}_{zz^\prime}^c\right)^2\vphantom{{\cal O}_{yy^\prime}^\odot}\right\}.
\eeqy

\subsection{Required Experimental Measurements in $\gamma N\to\pi\pi N$}

As in the case of $\pi N\to\pi\pi N$, we can examine which observables need to
be measured in order to extract information on the helicity or transversity
amplitudes. As there are eight such amplitudes, a minimum of eight measurements
must be made at each kinematic point (recall that these observables depend on 5 kinematic
variables) to obtain the absolute magnitudes of the
helicity or transversity amplitudes. In terms of our choice of transversity basis, these
measurements are the differential cross section, along with $P_y$, 
$P_{y^\prime}$, ${\cal O}_{yy^\prime}$, $I^\odot$, $P_y^\odot$, 
$P_{y^\prime}^\odot$ and ${\cal O}_{yy^\prime}^\odot$. 

The eight phases of the transversity amplitude mean that there are seven
independent phase differences that can be extracted, and seven measurements are
needed for this. For instance, the relative phases (in what should be an obvious
notation) $\phi_1^--\phi_2^-$, $\phi_1^+-\phi_2^+$, $\phi_3^--\phi_4^-$ and 
$ \phi_3^+-\phi_4^+$ require measurement of any four of the eight observables
$P_{x^\prime}$, $P_{z^\prime}$, ${\cal O}_{yx^\prime}$, ${\cal O}_{yz^\prime}$,
$P_{x^\prime}^\odot$, $P_{z^\prime}^\odot$, ${\cal O}_{yx^\prime}^\odot$ and
${\cal O}_{yz^\prime}^\odot$. $\phi_1^--\phi_3^-$ and $\phi_1^-+\phi_3^+$ may
then be extracted from measurement of any two observables from among $P_x$, $P_z$,
 ${\cal O}_{xy^\prime}$, ${\cal O}_{zy^\prime}$,
$P_x^\odot$, $P_z^\odot$, ${\cal O}_{xy^\prime}^\odot$ and
${\cal O}_{zy^\prime}^\odot$, along with use of the identities
$\phi_2^\pm-\phi_4^\pm=(\phi_2^\pm-\phi_1^\pm)+(\phi_1^\pm-\phi_3^\pm)+
(\phi_3^\pm-\phi_4^\pm)$. The remaining independent phase can then be extracted
from one of the observables that arise from linearly polarized photons. 
A `complete' set of experiments
will therefore require measurement of single, double and triple polarization observables,
in addition to the differential cross section.

\subsection{Parity conservation}

For the process $\gamma N\to\pi N$, parity conservation leads to the
relationships
\begin{equation}
{\cal M}_{-\lambda_N-\lambda_N^\prime}^{-\lambda_\gamma}(\theta)=
(-1)^{\lambda_\gamma-\lambda_N+\lambda_N^\prime}
{\cal M}_{\lambda_N\lambda_N^\prime}^{\lambda_\gamma}(\theta).
\end{equation}

The relationships that arise among the helicity amplitudes for $\gamma
N\to\pi\pi N$ are 
\begin{equation}
{\cal M}_{-\lambda_N-\lambda_N^\prime}^{-\lambda_\gamma}
(\theta,\theta_1,\phi_1)=(-1)^{\lambda_\gamma-\lambda_N+\lambda_N^\prime}
{\cal M}_{\lambda_N\lambda_N^\prime}^{\lambda_\gamma}
(\theta,\theta_1,2\pi-\phi_1).
\end{equation}
As was the case with $\pi N\to\pi\pi N$, these relations can not be used to 
decrease the number of independent helicity
amplitudes, but they can be used to determine which observables are even or odd under
the transformation $\phi_1\leftrightarrow 2\pi-\phi_1$.

\subsection{Construction of Transition Amplitudes}

\subsubsection{$\gamma N\to\pi N$}

In this case, there are two independent vectors $\vec{k}$, the momentum of the photon, and 
$\vec{q}$, the momentum of the pion. ${\cal A}$ must be an axial vector, while ${\cal
B}_{ij}$ must be a tensor. For real photons, $\vec{\varepsilon}\cdot\vec{k}=0$, so there can
be no $k_j$ terms in ${\cal B}_{ij}$. The forms that can be written are
\beqy
\vec{\cal A}&=& \alpha \hat{k}\times\hat{q},\nonumber\\
{\cal B}_{ij}&=& \beta_1 \delta_{ij}+\beta_2 \hat{k}_i \hat{q}_j +\beta_3 \hat{q}_i \hat{q}_j.
\eeqy
Comparing this with the form written by Chew, Goldberger, Low and Nambu \cite{cgln}
\beq
i{\cal M}=\chi^\dag\left(F_1 \vec{\sigma}\cdot\vec{\varepsilon}+iF_2\vec{\sigma}\cdot\hat{q}
\vec{\sigma}\cdot\hat{k}\times\vec{\varepsilon}+F_3\vec{\sigma}\cdot\hat{k}\hat{q}\cdot
\vec{\varepsilon}+F_4\vec{\sigma}\cdot\hat{q}\hat{q}\cdot\vec{\varepsilon}\right)\phi
\eeq
means that we can identify
\beqy
\alpha&=&i F_2,\,\,\,\,\,\,\,\,\,\,\,\,\,\,\,\,\,\,\,\,\,\beta_1=F_1-\hat{k}\cdot\hat{q}F_2,\nonumber\\
\beta_2&=&F_2+F_3,\,\,\,\,\,\,\,\,\beta_3=F_4.
\eeqy

From the explicit forms for $\hat{k}$, $\hat{q}$ and $\vec{\varepsilon}(\lambda)$, we can use
eqn (\ref{hell0}) to obtain ${\cal M}_1^\mp={\cal M}_4^\pm$, ${\cal M}_3^\mp=-{\cal M}_2^\pm$, (or, equivalently,
$b_4^\mp=b_1\pm$, $b_3^\mp=-b_2^\pm$)
leaving four independent helicity amplitudes, as expected. These helicity amplitudes are related to those of
Storrow \cite{storrow}, for example, by $N={\cal M}_2^+$, $S_1={\cal M}_4^+$, $S_2={\cal M}_1^+$ and 
$D={\cal M}_3^+$. Of the sixty-four observables,
thirty-two vanish identically. Furthermore, all thirteen remaining triple-polarization
observables are related to double- or single-polarization observables, or the differential
cross section, and three of the remaining fifteen double-polarization observables are related
to single-polarization observables, leaving a total of sixteen independent observables. The
remaining observables are given in terms of the helicity and transversity 
amplitudes in table \ref{tab:singlepolobs}

\begin{table}
\caption{Polarization observables of single pion photoproduction 
expressed as bilinear forms of the helicity amplitudes.\label{tab:singlepolobs}}
\begin{center}
\begin{tabular}{ll|l|l|l|ll}
\hline 
\hline 
\multicolumn{2}{c|}{Observable} & usual name
 & helicity form& transversity form & Measurements \\ \hline
$I_0$ & $\left(-{\cal O}^c_{yy^\prime}\right)$& $I_0$  & $ \left|{\cal M}_1^+\right|^2 + \left|{\cal 
M}_2^+\right|^2 + \left|{\cal M}_3^+\right|^2 + \left|{\cal 
M}_4^+\right|^2$ & 
$ \left|b_1^+\right|^2 + \left|b_2^+\right|^2 + \left|b_3^+
\right|^2 + \left|b_4^+\right|^2$ & $\left\{-;-;-\right\},\,\,\left\{L\left(\frac{\pi}{2},0\right);x;y^\prime\right\}$\\
$P_y$ & $\left(-P^c_{y^\prime}\right)$& $T$ & $ -2\Im\left({\cal M}_1^+{\cal M}_3^{+*} +
{\cal M}_2^+{\cal M}_4^{+*}\right)$ & 
 $ \left|b_1^+\right|^2 + \left|b_2^+\right|^2 - \left|b_3^+
\right|^2 - \left|b_4^+\right|^2$& $\left\{-;y;-\right\},\,\,\left\{L\left(\frac{\pi}{2},0\right);-;y^\prime\right\}$\\
$P_{y^\prime}$ & $\left(-P^c_y\right)$& $P$ & $ 2\Im\left({\cal M}_1^+{\cal M}_2^{+*}+
{\cal M}_3^+{\cal M}_4^{+*}\right)$ & 
$ \left|b_1^+\right|^2 - \left|b_2^+\right|^2 
+ \left|b_3^+\right|^2 - \left|b_4^+\right|^2$ & $\left\{-;-;y^\prime\right\},\,\,\left\{L\left(\frac{\pi}{2},0\right);y;-\right\}$ \\
${\cal O}_{xx^\prime}$ & $\left(-{\cal O}^c_{zz^\prime}\right)$& $T_x$ & $ 2\Re\left(-{\cal M}_2^+{\cal M}_3^{+*} - 
{\cal M}_1^+{\cal M}_4^{+*}\right)$ & 
 $ 2\Re\left(- b_2^+b_3^{+*} + 
  b_1^+b_4^{+*}\right)$ &  $\left\{-;x;x^\prime\right\},\,\,\left\{L\left(\frac{\pi}{2},0\right);z;z^\prime\right\}$ \\
${\cal O}_{xz^\prime}$ & $\left({\cal O}^c_{zx^\prime}\right)$& $T_z$ & $ 2\Re\left({\cal M}_1^+{\cal M}_3^{+*} - 
{\cal M}_2^+{\cal M}_4^{+*}\right)$ & 
$ -2\Im\left(- b_2^+b_3^{+*} - 
b_1^+b_4^{+*}\right)$ & $\left\{-;x;z^\prime\right\},\,\,\left\{L\left(\frac{\pi}{2},0\right);z;x^\prime\right\}$ \\
${\cal O}_{yy^\prime}$ & $\left(-I^c\right)$& $\Sigma$ & $ 2\Re\left(-{\cal M}_2^+{\cal M}_3^{+*} + 
{\cal M}_1^+{\cal M}_4^{+*}\right)$ & 
 $ \left|b_1^+\right|^2 - \left|b_2^+\right|
^2 - \left|b_3^+\right|^2 + \left|b_4^+\right|^2$ & $\left\{-;y;y^\prime\right\},\,\,\left\{L\left(\frac{\pi}{2},0\right);-;-\right\}$\\
${\cal O}_{zx^\prime}$ & $\left({\cal O}^c_{xz^\prime}\right)$& $L_x$ & $ 2\Re\left({\cal M}_1^+{\cal M}_2^{+*} - 
{\cal M}_3^+{\cal M}_4^{+*}\right)$ & 
$ -2\Im\left(- b_2^+b_3^{+*} +
 b_1^+b_4^{+*}\right)$ & $\left\{-;z;x^\prime\right\},\,\,\left\{L\left(\frac{\pi}{2},0\right);x;z^\prime\right\}$ \\
${\cal O}_{zz^\prime}$ & $\left(-{\cal O}^c_{xx^\prime}\right)$& $L_z$ & $ -\left|{\cal M}_1^+\right|^2 + \left|{\cal 
M}_2^+\right|^2 + \left|{\cal M}_3^+\right|^2 - \left|{\cal 
M}_4^+\right|^2$ & 
$2\Re\left(b_2^+b_3^{+*} +b_1^+b_4^{+*}\right)$ & $\left\{-;z;z^\prime\right\},\,\,\left\{L\left(\frac{\pi}{2},0\right);x;x^\prime\right\}$ \\
$P_x^\odot$ & $\left({\cal O}^s_{zy^\prime}\right)$& $F$ & $ 2\Re\left({\cal M}_1^+{\cal M}_3^{+*} + 
{\cal M}_2^+{\cal M}_4^{+*}\right)$ & 
 $ -2\Im\left(b_1^+b_3^{+*} + b_2^+b_4^{+*}\right)$ & $\left\{c;x;-\right\},\,\,\left\{L\left(\pm\frac{\pi}{4}\right);z;y^\prime\right\}$\\
$P_z^\odot$ & $\left(-{\cal O}^s_{xy^\prime}\right)$& $E$ & $ -\left|{\cal M}_1^+\right|^2 - \left|{\cal 
M}_2^+\right|^2 + \left|{\cal M}_3^+\right|^2 + \left|{\cal 
M}_4^+\right|^2$ & 
$ 2\Re\left(- b_1^+b_3^{+*} -  b_2^+b_4^{+*}\right)$ &  $\left\{c;z;-\right\},\,\,\left\{L\left(\pm\frac{\pi}{4}\right);x;y^\prime\right\}$\\
$P^\odot_{x^\prime}$ & $\left(-{\cal O}^s_{yz^\prime}\right)$& $C_x$ & $ -2\Re\left({\cal M}_1^+{\cal M}_2^{+*} + 
{\cal M}_3^+{\cal M}_4^{+*}\right)$ 
& $ -2\Im\left(- b_1^+b_2^{+*} - b_3^+b_4^{+*}\right)$ & $\left\{c;-;x^\prime\right\},\,\,\left\{L\left(\pm\frac{\pi}{4}\right);y;z^\prime\right\}$\\
$P^\odot_{z^\prime}$ & $\left({\cal O}^s_{yx^\prime}\right)$& $C_z$ & $ \left|{\cal M}_1^+\right|^2 - \left|{\cal 
M}_2^+\right|^2 + \left|{\cal M}_3^+\right|^2 - \left|{\cal 
M}_4^+\right|^2$ & 
$ 2\Re\left(-b_1^+b_2^{+*} - b_3^+b_4^{+*}\right)$ & $\left\{c;-;z^\prime\right\},\,\,\left\{L\left(\pm\frac{\pi}{4}\right);y;x^\prime\right\}$\\
$P_x^s$ & $\left(-{\cal O}^\odot_{zy^\prime}\right)$& $H$ & $ 2\Im\left({\cal M}_1^+{\cal M}_2^{+*} -
{\cal M}_3^+{\cal M}_4^{+*}\right)$ & 
$ 2\Re\left(b_1^+b_3^{+*} -b_2^+b_4^{+*}\right)$ & $\left\{L\left(\pm\frac{\pi}{4}\right);x;-\right\},\,\,\left\{c;z;y^\prime\right\}$\\
$P_z^s$ & $\left({\cal O}^\odot_{xy^\prime}\right)$& $G$ & $ 2\Im\left(-{\cal M}_2^+{\cal M}_3^{+*} + 
{\cal M}_1^+{\cal M}_4^{+*}\right)$ & 
$ -2\Im\left(b_1^+b_3^{+*} - b_2^+b_4^{+*}\right)$ & $\left\{L\left(\pm\frac{\pi}{4}\right);z;-\right\},\,\,\left\{c;x;y^\prime\right\}$\\
$P^s_{x^\prime}$ & $\left({\cal O}^\odot_{yz^\prime}\right)$& $O_x$ & $ -2\Im\left({\cal M}_1^+{\cal M}_3^{+*} - 
{\cal M}_2^+{\cal M}_4^{+*}\right)$ & 
$ 2\Re\left(-b_1^+b_2^{+*} + b_3^+b_4^{+*}\right)$ & $\left\{L\left(\pm\frac{\pi}{4}\right);-;x^\prime\right\},\,\,\left\{c;y;z^\prime\right\}$\\
$P^s_{z^\prime}$ & $\left(-{\cal O}^\odot_{yx^\prime}\right)$& $O_z$ & $ -2\Im\left({\cal M}_2^+{\cal M}_3^{+*} + 
{\cal M}_1^+{\cal M}_4^{+*}\right)$ & 
$ -2\Im\left(b_1^+b_2^{+*} - b_3^+b_4^{+*}\right)$ & $\left\{L\left(\pm\frac{\pi}{4}\right);-;z^\prime\right\},\,\,\left\{c;y;x^\prime\right\}$\\
\hline 
\end{tabular}
\end{center} 
\end{table}
The relationships among these observables, obtained from consideration of the
transversity amplitudes, are
\beqy
&&\left(P_{x^\prime}^\odot\mp P^s_{z^\prime}\right)^2
+\left(P_{z^\prime}^\odot\pm P^s_{x^\prime}\right)^2=\left(1\pm P_y\right)^2-
\left(P_{y^\prime}\pm {\cal O}_{yy^\prime}\right)^2,\nonumber\\
&&\left(P_x^\odot\pm P^s_z\right)^2+\left(P_z^\odot\mp P^s_x\right)^2=
\left(1\pm P_{y^\prime}\right)^2-
\left(P_y\pm {\cal O}_{yy^\prime}\right)^2,\nonumber\\
&&\left({\cal O}_{xz^\prime}\mp{\cal O}_{zx^\prime}\right)^2
+\left({\cal O}_{xx^\prime}\pm{\cal O}_{zz^\prime}\right)^2
=\left(1\pm{\cal O}_{yy^\prime}\right)^2-
\left(P_y\pm P_{y^\prime}\right)^2.
\eeqy
These lead to the inequalities
\beqy
&&\left|1\pm P_y\right|\ge\left\{\left|P_{x^\prime}^\odot\mp P^s_{z^\prime}\right|
,\,\,\,\,\left|P_{z^\prime}^\odot\pm P^s_{x^\prime}\right|,\,\,\,\,
\left|P_{y^\prime}\pm {\cal O}_{yy^\prime}\right|\right\},\nonumber\\
&&\left|1\pm P_{y^\prime}\right|\ge\left\{\left|P_x^\odot\pm
P^s_z\right|,\,\,\,\,\left|P_z^\odot\mp P^s_x\right|,\,\,\,\,
\left|P_y\pm {\cal O}_{yy^\prime}\right|\right\},\nonumber\\
&&\left|1\pm{\cal O}_{yy^\prime}\right|\ge\left\{\left|{\cal O}_{xz^\prime}\mp{\cal O}_{zx^\prime}\right|
,\,\,\,\,\left|{\cal O}_{xx^\prime}\pm{\cal O}_{zz^\prime}\right|,\,\,\,\,
\left|P_y\pm P_{y^\prime}\right|\right\},
\eeqy
and
\beqy
&&1+ P_y^2\ge\left\{\left(P_{x^\prime}^\odot\right)^2+ \left(P^s_{z^\prime}\right)^2
,\,\,\,\,\left(P_{z^\prime}^\odot\right)^2+ \left(P^s_{x^\prime}\right)^2,\,\,\,\,
P_{y^\prime}^2+ {\cal O}_{yy^\prime}^2\right\},\nonumber\\
&&1+ P_{y^\prime}^2\ge\left\{\left(P_x^\odot\right)^2+
\left(P^s_z\right)^2,\,\,\,\,\left(P_z^\odot\right)^2+ \left(P^s_x\right)^2,\,\,\,\,
P_y^2+ {\cal O}_{yy^\prime}^2\right\},\nonumber\\
&&1+{\cal O}_{yy^\prime}^2\ge\left\{{\cal O}_{xz^\prime}^2+{\cal O}_{zx^\prime}^2
,\,\,\,\,{\cal O}_{xx^\prime}^2+{\cal O}_{zz^\prime}^2,\,\,\,\,
P_y^2+ P_{y^\prime}^2\right\}.
\eeqy

\subsubsection{$\gamma N\to\pi\pi N$}

\newcommand{\uk}{\hat{k}}
\newcommand{\uq}{\hat{q}_1} 
\newcommand{\uQ}{\hat{p}_2}
\newcommand{\un}{\hat{n}}
\newcommand{\sig}{\sigma}

For this process, we have three independent vectors, $\hat{k}$, $\hat{p}_2$ and $\hat{q}_1$ with
which to construct a vector for $\vec{\cal A}$ and a pseudotensor for ${\cal B}_{ij}$.
However, using these leads to the difficulty that there are too many structures left in ${\cal
B}_{ij}$. To avoid this problem, we define an axial vector $\un$ as 
$\un=\uk \times \uQ/N$, with $N=|\uk \times \uQ|=\sin\theta$. $\un$ defines the
$y$ axis while the $x$ axis is defined by $\un \times \uk=(\uQ-\uk \uk \cdot 
\uQ)/N$. We can now write
\begin{equation}
\uq=\uq \cdot \uk \uk + \uq \cdot \un \un + 
(\uq \cdot \uQ - \uq \cdot \uk \uk \cdot \uQ) (\uQ-\uk \uk \cdot \uQ)/N^2
\end{equation}
and use the axial vector $\un$ and the pseudoscalar ${\cal P}=\uq \cdot \uk \times \uQ=N \uq \cdot \un$
instead of $\uq$ to build the structures that make up $A^i$ and $B^{ij}$. $\un$ and ${\cal P}$ can
appear only once in these structures since, by expanding the product of Levi-Civita 
tensors, it is easy to show that ${\cal P}^2$ is a scalar that depends only on quantities
already defined 
\begin{equation}
{\cal P}^2=1-\left(\uk \cdot \uQ\right)^2-\left(\uk \cdot \uq\right)^2-\left(\uQ \cdot \uq\right)^2-
      2 \uk \cdot \uQ   \uk \cdot \uq   \uQ \cdot \uq,
\end{equation}
while $\un^i \un^j$ can be expressed as
\begin{equation} \label{eq:ninj}
\un^i \un^j=\delta^{ij} 
           - \frac{\uQ^i \uQ^j +\uk^i \uk^j + \uk \cdot \uQ ( \uk^i \uQ^j + \uQ^i \uk^j)}{N^2}.
\end{equation}
The vectors that can make up $A^i$ are $k^i, q_1^i$ and ${\cal P}\un^i$. Since $\varepsilon \cdot \uk=0$, 
only two structures remain.
Similarly $B^{ij}$ can be expressed as a sum of 
$\un^i \uQ^j$, $\un^i \uk^j$, 
$\uQ^i \un^j$, ${\cal P} \uQ^i \uQ^j$, ${\cal P} \uQ^i \uk^j$, 
${\cal P} \delta^{ij}$, 
$\epsilon^{ijk} \uk^k$, $\epsilon^{ijk} \uQ^k$ and ${\cal P} \epsilon^{ijk} \un^k$.

Expressing $\varepsilon^i$ as 
\begin{eqnarray}
\varepsilon^i&=&\varepsilon \cdot \uk \uk^i + \varepsilon \cdot \un \un^i + 
(\varepsilon \cdot \uQ - \varepsilon \cdot \uk \uk \cdot \uQ) (\uQ^i-\uk^i \uk \cdot \uQ)/N^2\\
&=&\varepsilon \cdot \un \un^i + \varepsilon \cdot \uQ (\uQ^i-\uk^i \uk \cdot \uQ)/N^2,
\end{eqnarray}
it is easy to show that the last three structures can be expressed as
\begin{eqnarray}
   \epsilon^{ijk} \uk^k \varepsilon^i&=&\frac{\un^i \uQ^j               - \un^i \uk^j \uQ \cdot \uk - \uQ^i
   \un^j              }{N}\varepsilon^i\\
   \epsilon^{ijk} \uQ^k \varepsilon^i&=&\frac{\un^i \uQ^j \uk \cdot \uQ - \un^i \uk^j               + \uQ^i
   \un^j \uk \cdot \uQ}{N}\varepsilon^i\\
{\cal P}\epsilon^{ijk} \un^k \varepsilon^i&=&{\cal P}\frac{\uQ^i \uk^j}{N}\varepsilon^i.
\end{eqnarray}
These three structures can therefore be omitted from the construction of the amplitude. Finally, we
write
\begin{eqnarray}
\vec{{\cal A}}&=& \alpha_1 \uq+\alpha_2 {\cal P}\un,\nonumber\\
{\cal B}^{ij}&=&\beta_1\un^i\uQ^j+\beta_2\un^i \uk^j+\beta_3\uQ^i \un^j
+\beta_4 {\cal P} \uQ^i \uQ^j+\beta_5 {\cal P} \uQ^i \uk^j+\beta_6{\cal P} \delta^{ij}.
\end{eqnarray}

As discussed previously, parity conservation can be used to tell which
observables are even and which are odd under the transformation
$\phi_1\leftrightarrow 2\pi-\phi_1$. In the previous subsection, we listed
the non-vanishing observables for $\gamma n\to N\pi$. The corresponding
observables in $\gamma n\to N\pi\pi$ are all even under the transformation
in $\phi_1$. The variables that vanish in $\gamma N\to\pi N$, are non-zero for
$\gamma N\to\pi\pi N$, but are odd under the $\phi_1$ transformation.

\section{Conclusion and Outlook}

We have developed a set of polarization observables that are applicable to final
states that contain two pseudoscalar mesons and a spin-1/2 baryon, such as
$N\pi\pi$, and have examined the observables that arise using both photon and
pion (or other pseudoscalar meson) beams. We have written these observables in
terms of both helicity and transversity amplitudes, obtained relationships among
them, and used these to list inequalities that these observables satisfy. We
have also indicated the measurements that are needed for each observable.
The framework that we have used is a very simple one: undoubtedly, the
expressions for the observables and the relationships among them can be derived
in a more elegant manner. 

While we have discussed helicity and one set of transversity amplitudes, there
remains the possibility of defining yet another set of amplitudes, and writing
the observables in terms of these. In the com frame, the momenta of the final
particles satisfy $\vec{p}_2+\vec{q}_1+\vec{q}_2=\vec{0}$, which means that
they define a plane. The normal to this plane can be defined by
$\vec{p}_2\times\vec{q}_1$, and this offers another natural axis for
quantization of the spin of the final nucleon. One possible advantage of using
this axis is that it automatically incorporates information about the entire final
state, not just the final nucleon. Whether this leads to any particular
advantage, simplification or insight into the observables, the relationships among them, or even
in the amplitudes themselves, awaits exploration.

As we stated at the start of this manuscript, polarization observables are
crucial for extracting resonance information from scattering data. Differential
cross sections, presented in whatever form, will only provide information on the
magnitudes of helicity or transversity amplitudes. Phase information is crucial,
and this is only available from measurement of a number of different
observables. This is well known for processes like $\gamma N\to\pi N$. The same
is true, or perhaps, even more true, for processes like $\gamma N\to\pi\pi N$.
Models with quite different input can and will succeed in describing the total
and differential cross section, but the polarization observables will serve to
distinguish among such models.

A number of these observables can be measured in the near future at existing facilities, for
a number of processes. Indeed, the photon polarization  asymmetry $I^\odot$, has already been
measured at Jefferson Laboratory \cite{strauch} for $\gamma p\to p\pi^+\pi^-$, and the
analysis is continuing at present. Clearly, this variable can be measured in other channels,
and there are plans to do so for $\gamma p\to p\pi^0\pi^0$ at Bonn \cite{thoma}. The
existence of polarized targets means that $P_x$, $P_y$ and $P_z$ are accessible, and coupling
such targets with circularly polarized beams allows measurement of $P_x^\odot$, $P_y^\odot$
and $P_z^\odot$. The use of linearly polarized photons opens the door to measurements of
$P_x^{s,c}$, $P_y^{s,c}$, $P_z^{s,c}$ and  $I^{s,c}$. For processes with a hyperon in the
final state, such as $\gamma N\to\pi K\Lambda$, the self-analyzing decay of the hyperon
allows its polarization to be determined, in principle allowing many more observables to be
measured, including a number of triple-polarization ones. For processes like $\pi N\to\pi\pi
N$, three of the observables are readily available with polarized targets. All others require
the measurement of recoil polarization. Unfortunately, there are at present no existing
hadronic beams facilities that will allow us to capitalize on these observables.

We have not attempted to explore the properties of the observables that we
described herein, apart from a brief discussion of the oddness or evenness under
the $\phi_1$ transformation. In particular, we have said nothing on their values
at special values of $\theta$, for instance, such as $\theta=0$ or $\pi$. This
is left for a possible future manuscript. In the near future, we plan to explore
a number of these observables in the framework of an existing model for the
photoproduction of two pseudoscalar mesons off a nucleon target. In particular,
the sensitivity of the observables to the details of the underlying dynamics, as
well as the rich structure of these observables, will be discussed.

\section*{Acknowledgment}

WR thanks B. Mecking for reading the manuscript, and F. Gross for discussions.
This work was supported by the Department of Energy through contract
DE-AC05-84ER40150, under which the Southeastern Universities Research
Association (SURA) operates the Thomas Jefferson National Accelerator Facility
(TJNAF).

\end{document}